# On-Chip and Off-Chip TIA Amplifiers for Nanopore Signal Readout: Design, Performance and Challenges

K. Ashoka Deepthi[1a]      Manoj Varma[2b]      Arup Polley[1c]

[1]Department of Electronic System Engineering (DESE), IISc Bengaluru

[2] Centre for Nano Science and Engineering (CeNSE), IISc Bengaluru

e-mail: [a]deepthia@iisc.ac.in      [b]mvarma@iisc.ac.in      [c]aruppolley@iisc.ac.in

**Abstract**

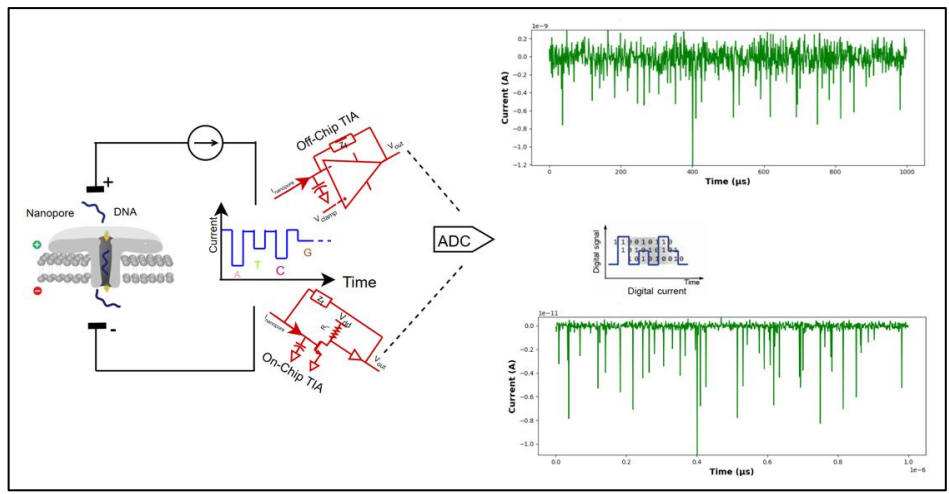

Advancements in biomedical research have driven continuous innovations in sensing and diagnostic technologies. Among these, nanopore based single molecule sensing and sequencing is rapidly emerging as a powerful and versatile sensing methodology. Advancements in nanopore based approaches require concomitant improvements in the electronic readout methods employed, from the point of low noise, bandwidth and form factor. This article focuses on current sensing circuits designed and employed for ultra-low noise nanopore signal readout, addressing the fundamental limitations of traditional off chip transimpedance amplifiers (TIAs), which suffer from high input parasitic capacitance, bandwidth constraints, and increased noise at high frequencies. This review explores the latest design schemes and circuit structures classified into on-chip and off-chip TIA designs, highlighting their design implementation, performance, respective challenges and explores the interplay between noise performance, capacitance, and bandwidth across diverse transimpedance amplifier (TIA) configurations. Emphasis is placed on characterizing noise response under varying parasitic capacitance and operational frequencies, a systematic evaluation not extensively addressed in prior literature while also considering the allowable input current compliance range limitations. The review also compares the widely used Axopatch 200B system to the designs reported in literature. The findings offer valuable insights into optimizing TIA designs for enhanced signal integrity in high speed and high sensitivity applications focusing on noise reduction, impedance matching, DC blocking, and offset cancellation techniques.

## 1. INTRODUCTION

Nanopore based biosensors utilize nanoscale apertures to identify and analyze individual biomolecules such as DNA [1] and proteins [2]. They work by monitoring changes in ionic current during the passage of single molecules through the nanopores situated in an electrolyte filled medium. Distinct signals are generated based on molecular characteristics such as size, shape, and surface charges [3]. This precise detection capability has led nanopore sensors to be widely employed in critical applications such as single molecule DNA sequencing [4], protein identification [5], real-time single molecule sensing [6], and peptide analysis [7]. Accurate measurements at picoampere (pA) current levels are essential for these applications, demanding specialized and highly sensitive amplification circuitry. Transimpedance amplifiers (TIAs) play a crucial role in nanopore sensing, enabling accurate detection of picoampere level ionic currents [8]. These amplifiers convert current into voltage using feedback elements, which can be categorized into resistive feedback and capacitive feedback configurations. In resistive feedback, transimpedance gain is determined by the resistance value, often requiring high value feedback resistors in the GΩ range to minimize noise. However, integrating such large resistors into compact semiconductor chips introduces parasitic capacitances that can limit the measurement bandwidth. Several advanced TIA designs have been reported to mitigate performance limiting effects such as parasitic capacitance, noise, and offsets to enhance detection accuracy in nanopore applications. This review provides a structured introduction to this area covering the fundamental concepts as well as the salient progress made in nanopore signal readout. This article begins with an introduction to the fundamental concepts in nanopore signal measurement and then covers the basics of biological and solid state nanopores, highlighting their operational principles and their significance in biosensing. The third section describes the experimental setup employed in nanopore sensing, including the principles behind ionic current generation, measurement, and characterization. A unique switch based equivalent circuit model is proposed in this study to evaluate the performance of amplifiers during nanopore simulations. The fourth section extends this analysis by focusing on electronic readout methods, specifically examining various TIA architectures. This section discusses the evolution from basic TIA configurations to advanced designs incorporating DC stabilization [13] and sophisticated noise reduction strategies [35], such as capacitive feedback, active feedback loops, and auto zeroing techniques [11][12]. The TIAs discussed in this paper are systematically classified into two categories: on-chip implementations, which are typically based on CMOS technology, and off-chip implementations, which involve discrete components or are realized at the board level. Within the on-chip and off-chip category, different TIA implementations are compared based on key metrics such as gain, bandwidth, supply requirements, noise performance, number of sensing channels, parasitic capacitance, power consumption, and current range. Similarly, a comparative analysis is conducted between on-chip and off-chip implementations using these performance metrics. By systematically examining these approaches and highlighting the design methodology, performance, and challenges of each model, this review offers valuable insights into the trade-offs between on-chip and off-chip TIA designs. The review also uniquely evaluates noise variation by systematically varying input parasitic capacitance and frequency, an aspect not extensively explored in previous studies. It further compares the widely used Axopatch 200B off-chip amplifier, which despite having high input parasitic capacitance, achieves low noise performance and less leakages due to internal cooling techniques and specialized circuit design, with other off-chip TIA configurations. This comparative analysis highlights the advantages and limitations of each approach, offering guidance for circuit designers in selecting optimal TIA solutions for high precision nanopore sensing applications.

## 2. NANOPORE TECHNOLOGIES: BIOLOGICAL AND SOLID-STATE APPROACHES

### 2.1. Biological Nanopores

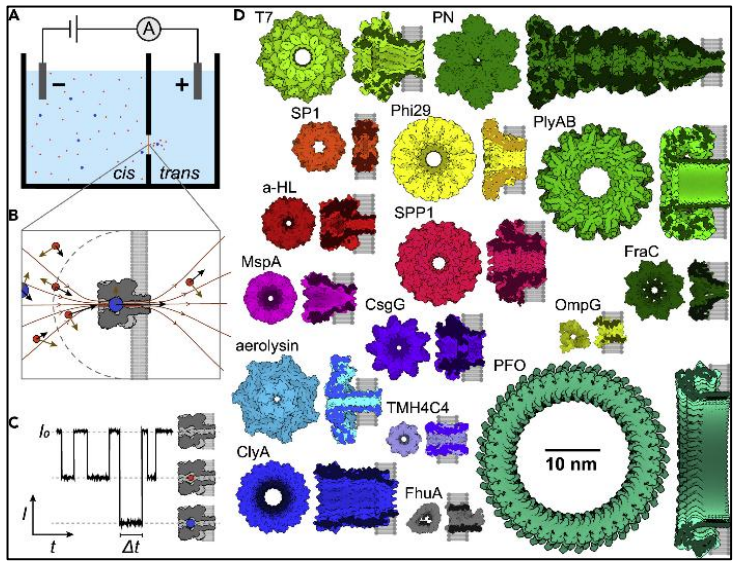

*Figure 1: Biological Nanopore-Based Sensing*

*(A) Diagram of a nanopore detection system containing two distinct molecular analytes (red and blue) dispersed in an electrolyte medium (light blue). These analytes pass through a biological nanopore situated within a lipid membrane (orange). (B) Illustration showing the effective capture zone (dotted circle), distribution of electric field lines (gray), Brownian motion (brown arrows), and electrophoretic forces (black arrows), where arrow length indicates relative strength. (C) Representative ionic current signal during nanopore measurements, highlighting the open pore baseline current ($I_0$), transient reductions in current (blockades), and the corresponding translocation durations (Δt) for two analytes of varying sizes. (D) Structures of 16 biological nanopores employed in nanopore sensing studies (excluding TMH4C4). Shown from top left to bottom right: T7 (light green, PDB: 3J4A), modified proteasome pore (dark green), SP1 (orange, 1TR0), Phi29 (yellow, 1JNB), PlyAB (electric green, 4V2T), α-hemolysin (red, 7AHL), SPP1 (magenta, 2JES), OmpG (yellow-green, 2JQY), FraC (forest green, 4TSY), MspA (pink, 1UUN), CsgG (purple, 4UV3), PFO (turquoise, 2BK1), aerolysin (cyan, 5JZT), TMH4C4 (lavender, 6M6Z), engineered FhuA ΔC/ΔD4L (gray, 1BY3), and 13-subunit ClyA (blue, 6MRU). Reproduced with permission from reference [18]. Copyright (2022) Elsevier.*

The concept of utilizing nanopores for molecular detection was first proposed by David Deamer in 1989. However, it gained significant attention after Kasianowicz et al. demonstrated DNA translocation through α-hemolysin (α-HL) in 1996 [61] . This breakthrough was inspired by the natural transport mechanisms of biomolecules across cell membranes via channel proteins, laying the foundation for modern nanopore technologies. In the early stages of nanopore research, most nanopores were derived from intrinsic channel proteins, such as α-hemolysin (α -HL) [62] , Mycobacterium smegmatis porin A (MspA) [63], and the bacteriophage phi29 DNA packaging motor[64]. These biological nanopores provided essential insights into single molecule sensing applications. Mayer et al. [54] provided a comprehensive review summarizing sixteen biological nanopores, detailing their properties and potential applications. Recent developments have focused on engineering biological nanopores to enhance their functionality and broaden their applications for protein engineering for enhanced selectivity and sensitivity [13], hybrid nanopore systems [14], and expansion beyond DNA sequencing [15]. Biological nanopores offer a significant advantage over solid-state nanopores fabricated through conventional manufacturing techniques, primarily due to their high reproducibility and precisely defined atomic level structures [17]. As a result, commercially available nanopore based DNA sequencing technologies predominantly utilize arrays of biological nanopores. One of the key strengths of biological nanopores is their versatile pore lumen sizes, which can range from a few angstroms to several nanometers. This variability enables their adaptation for diverse biosensing and molecular detection applications, offering high specificity for target molecules [65][66]. However, despite their reproducibility and functional adaptability, biological nanopores exhibit certain limitations. They often lack the necessary chemical and mechanical stability required for some advanced sensing applications [16]. This limitation presents challenges in environments demanding extended durability or harsh chemical conditions, highlighting the need for further optimization or hybrid approaches integrating biological and synthetic nanopores.

## 2.2. Solid-state nanopore

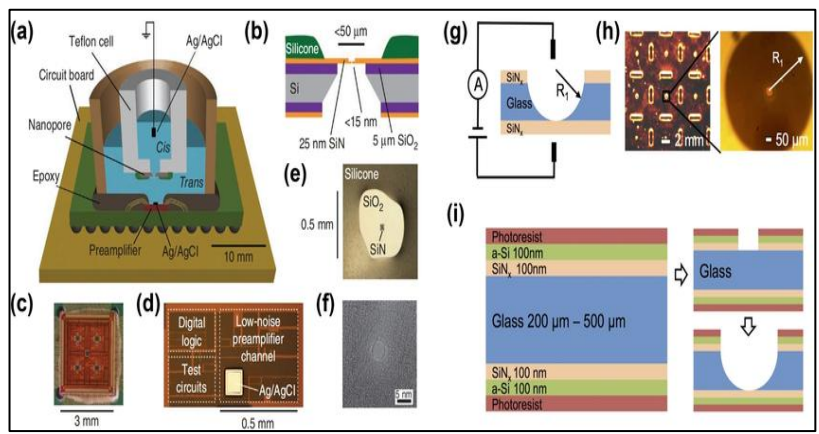

*Figure 2: Solid-State Nanopore Sensing Using Microfabricated Platform*
*(a) Conceptual diagram of the solid-state nanopore measurement system. (b) Side-view representation of a chip incorporating a low-dielectric thin-film membrane to minimize parasitic capacitance. (c) Optical image of a custom-designed 8-channel CMOS-based transimpedance amplifier array for current detection. (d) Enlarged view highlighting the architecture of an individual amplification channel. (e) Photograph of a silicon nitride membrane chip housed within a fluidic measurement chamber. (f) High-resolution transmission electron microscopy (TEM) image showing a 4 nm nanopore drilled into a silicon nitride membrane. (g–i) Schematic of a glass-based device featuring a suspended membrane that separates two ionic reservoirs, where a voltage differential is applied to drive ion flow through the nanopore. Reproduced with permission from reference [31]. ©2012 American Chemical Society.*

Solid-state nanopore fabrication technology was first demonstrated by Jiali Li and her collaborators [55] in 2001, marking the beginning of extensive research in this field. Over time, it has become a focal point of study due to its high mechanical stability, tunable pore dimensions, and potential for integration with electronic systems. Based on the fabrication mechanism, solid-state nanopore manufacturing can be categorized into two main approaches: top-down etching and bottom-up shrinkage techniques. The top-down approach involves direct pore formation using methods such as focused ion beam (FIB) etching [28] and high-energy electron beam drilling [29], plasma etching, and wet etching [30] enabling precise control over nanopore size and shape. In contrast, the bottom-up approach builds upon top-down techniques by employing methods like electron beam-assisted deposition [57] and atomic layer deposition (ALD) [56] to further refine nanopore characteristics. These techniques have facilitated the development of solid-state nanopores with adjustable diameters and controlled channel lengths, making them highly adaptable for various sensing applications. Among the materials used in fabrication, silicon [20], silicon nitride [21] and silicon oxide ,polymers[22],aluminum oxide[23] are widely employed due to their excellent stability and performance. Additionally, advanced two-dimensional materials such as graphene [24] and molybdenum disulfide [26], and MXene [25], phosphorene [27] have emerged as promising alternatives, offering enhanced electrical conductivity, molecular detection sensitivity, and improved structural properties.

## 3. NANOPORE CURRENT MEASUREMENT: PRINCIPLES, CIRCUIT MODELING, AND AMPLIFIER EVALUATION

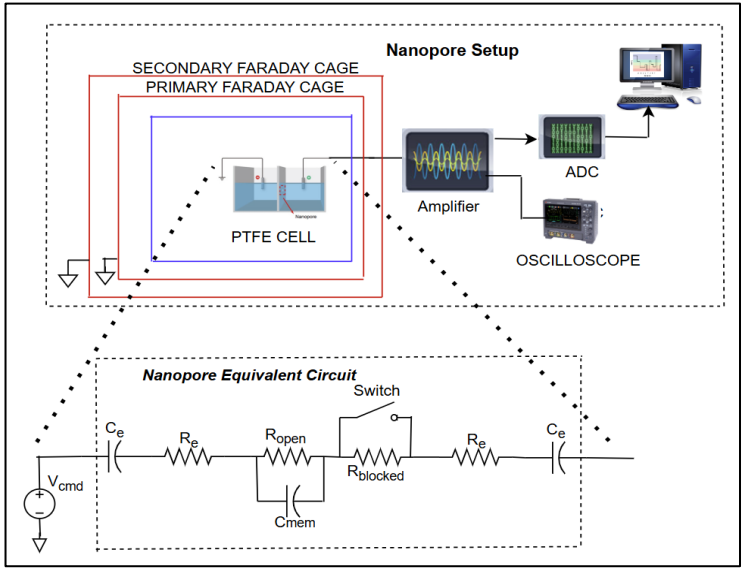

*Figure 3: Nanopore current measurement setup and representation of Equivalent Switch-Based Nanopore Circuit Model*

Nanopore current measurement involves a chip with a nanopore or nanopore array mounted on a PTFE chamber, separating two compartments (cis and trans) filled with electrolyte solutions [58]. Ag/AgCl electrodes establish a potential difference across the membrane, driving ionic current through the nanopore. To ensure precise measurements and minimize electromagnetic interference, the apparatus is enclosed within primary and secondary Faraday cages and placed on an anti-vibration table. Ionic currents are recorded using a low current measurement circuit, and analog signals are converted into digital data by an ADC, which also manages the amplifier's command voltage [60]. An oscilloscope provides real time monitoring, ensuring accurate signal capture. In nanopore based sequencing setups, DNA or RNA molecules introduced into the negatively biased compartment migrate electrophoretically through the nanopore [59]. As each nucleotide passes through, it generates a unique ionic current blockade with a specific dwell time. Analyzing these ionic current variations enables precise nucleotide identification, forming the theoretical basis for de novo DNA sequencing, to effectively represent the electrical behavior observed during nanopore measurements and sequencing, a uniquely defined equivalent circuit model is presented in this paper. This model incorporates several critical components: membrane capacitance ($C_{mem}$), typically ranging from 2 pF to 20 pF, and nanopore resistances categorized as open pore resistance ($R_{open}$) and blocked pore resistance ($R_{blocked}$), each usually in the range of several GΩ. Additionally, electrode resistance ($R_e$) and electrode capacitance ($C_e$) are included to complete the electrical characterization. Within this equivalent circuit representation, a switch symbolizes the two operational states of the nanopore open and closed operation. In the absence of analytes, the switch is short circuited, and the open pore resistance ($R_{open}$) combined in series with electrode resistance ($R_e$) defines the circuit, producing the baseline or open pore current. When an analyte enters the nanopore, the switch opens, introducing additional resistance. Consequently, the equivalent resistance transitions to blocked pore resistance ($R_{blocked}$), which in combination with membrane capacitance ($C_{mem}$), creates a parallel arrangement. This parallel assembly connects in series with electrode resistances ($R_e$), resulting in a blockade current. Furthermore, the controlled switching operation, driven by voltage pulses of varying durations, serves as a practical method to assess and verify the amplifier's gain and bandwidth performance.

## 4.  ELECTRONICS FOR NANOPORE READOUT

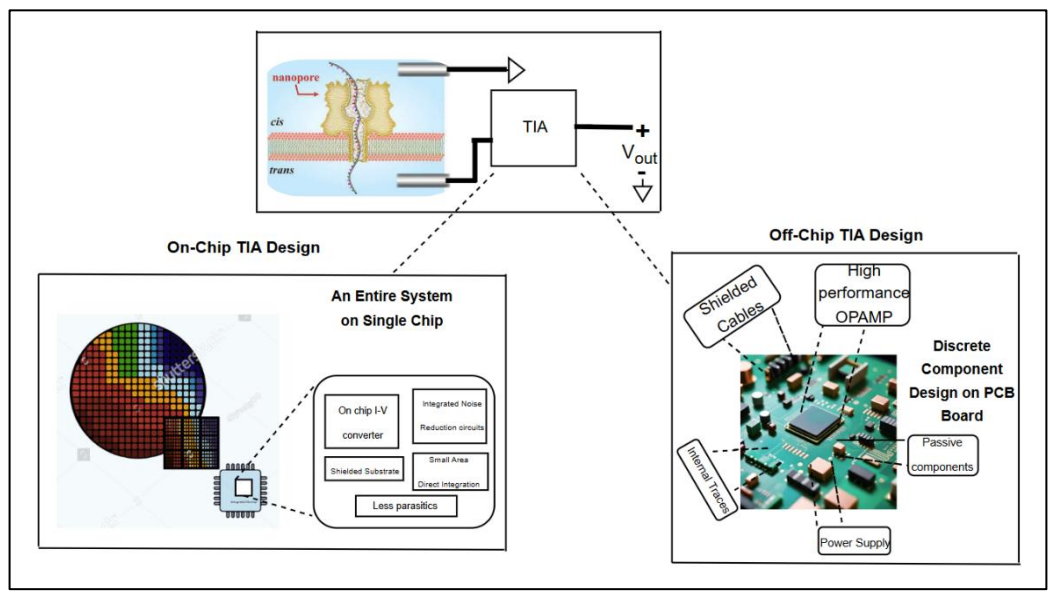

*Figure 4: Comparison Between On-Chip and Off-Chip Implementations*

To accurately analyze the ionic currents generated in nanopore sensing, there are two primary approaches for implementing amplifiers in nanopore readout systems: off-chip and on-chip methods. First, we discuss off-chip amplifier designs, which often use Operational Amplifiers (Op-Amps). These provide greater flexibility and ease of customization, making them suitable for research and prototyping. However, at nanopore range operating frequencies, off-chip implementations often experience increased leakage current and pronounced parasitic effects, as discussed in detail in the subsequent sections. Next, we discuss on-chip designs, typically utilizing Application-Specific Integrated Circuits (ASICs) or CMOS technology for TIA implementation, which provide high sensitivity and low noise due to their integrated architecture. These on-chip solutions are particularly advantageous for compact, high throughput applications, as they mitigate parasitic issues and ensure stable performance even at higher operational frequencies. This integration also lowers power consumption and improves reliability by eliminating external interconnects and stray capacitance. The choice between off-chip and on-chip approaches depends on the specific requirements for accuracy, scalability, and cost effectiveness in the nanopore sensing system. Table I provides a detailed comparison between these two methods, highlighting key aspects such as noise performance, signal integrity, power efficiency, design complexity, scalability, cost implications, and application suitability.

| Aspect | Off-Chip Method | On-Chip Method |
|---|---|---|
| Location and Integration | Discrete components placed on a printed circuit board, connected via wiring. | Integrated directly on the nanopore sensor chip, minimizing interconnect length. |
| Noise Performance | Higher noise susceptibility from long interconnects, external interference. | Low noise due to reduced parasitic capacitance and inductance, ensuring high signal to noise ratio (SNR). |
| Signal Integrity | Signal degradation due to RC effects in traces and electromagnetic interference. | High signal integrity with minimal signal degradation, maintaining accurate ionic current measurements. |
| Power Efficiency | Higher power consumption due to discrete component usage. | Lower power consumption due to optimized CMOS technology, suitable for portable devices. |
| Performance | Moderate performance due to external connections, higher parasitic effects, and increased noise at high frequencies. | High performance with lower noise, reduced parasitic, and stable operation at high frequencies due to integrated architecture. |
| Design Complexity | Simpler design requires a careful PCB layout to minimize noise and signal loss. | More complex design with challenges in noise isolation, thermal management, and power efficiency. |
| Scalability | Limited scalability due to space constraints and wiring complexity. | Highly scalable, enabling integration of multi nanopore arrays for high throughput applications. |
| Cost Implications | Lower initial cost with off the shelf components, ideal for small scale testing and prototyping. | Higher initial cost for custom CMOS fabrication but cost effective for mass production. |
| Application Suitability | Suitable for research, development, and proof of concept testing due to ease of modification and cost effectiveness. | Ideal for commercial, high-performance applications like DNA sequencing, biosensors, and portable diagnostic tools. |

*Table I: Comparison Between Off-Chip and On-Chip Amplifier Methods for Nanopore Based Sensing Systems*

## 4.1. Off-Chip Transimpedance Amplifier for Nanopore Based Systems

### 4.1.1. The Classical Transimpedance Amplifier

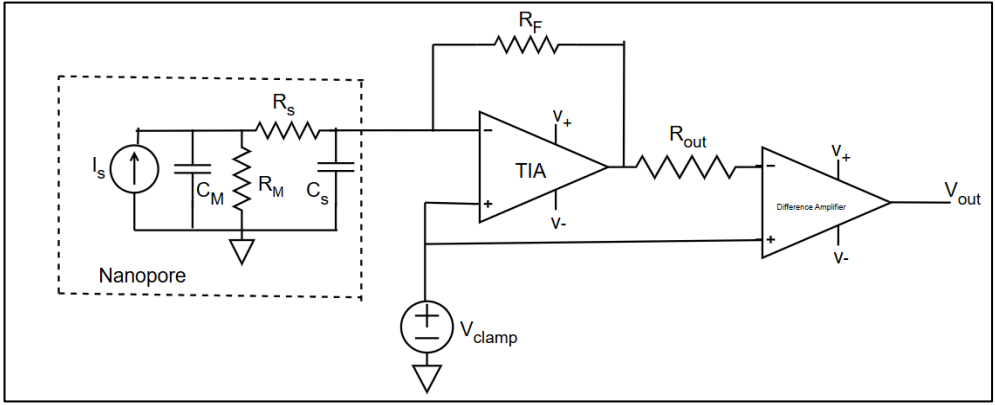

*Figure 5: Schematic of the Two-Stage, Low Noise Transimpedance Amplifier.*

**4.1.1.1 Design and Implementation**

The Classical Transimpedance Amplifier (TIA) [32] consists of a high value feedback resistor ($R_F$) in the first stage, which converts the input current into a voltage signal. A differential amplifier follows to eliminate any superimposed clamp voltage ($V_{clamp}$). This design is commonly used for basic current sensing, however, the presence of the $R_F\ C_F$ network introduces a low pass filtering effect, where the cutoff frequency is given by:

$$f_c = \frac{1}{2*\pi*R_F*C_F}$$

As the detection current decreases, the required feedback resistance ($R_F$) increases significantly, often reaching the GΩ level. For instance, a 1GΩ feedback resistor introduces at least 300fF of stray capacitance, which severely limits the bandwidth. Furthermore, each sensing channel requires a dedicated amplifier, increasing circuit complexity, reducing scalability, and leading to higher power consumption.

**4.1.1.2 Performance**

The bandwidth of the classical TIA is significantly constrained by the $R_F\ C_F$ network, and when $R_F$ reaches the GΩ range, the system becomes limited to approximately 530 Hz. Such a bandwidth is insufficient for high-speed applications like nanopore sensing, which requires bandwidths in the kHz to MHz range. Moreover, large $R_F$ values introduce significant thermal noise, while the differential amplifier further contributes to noise, resulting in a poor signal to noise ratio (SNR). These issues are particularly problematic in applications where low noise performance is essential. Additionally, large $R_F$ values impact the phase margin, potentially causing peaking or oscillations, necessitating compensation techniques such as pole zero cancellation networks.

**4.1.1.3 Challenges**

At ultra low detection currents in the picoampere or femtoampere range, the classical TIA suffers from non-linearity due to leakage currents from the board, offset voltages, and input bias currents of the operational amplifier, all of which degrade measurement accuracy. The reliance on a high value $R_F$ also makes the design susceptible to noise and stability issues, particularly in precision applications where accuracy is critical. Furthermore, the requirement for a dedicated amplifier per channel limits scalability, increasing system complexity and power consumption. While the classical TIA remains effective for basic current sensing, its severe bandwidth limitations, high noise levels, and lack of scalability make it impractical for high speed and high sensitivity applications. TIA also exhibits sensitivity to temperature variations, as fluctuations in $R_F$ lead to gain drift and instability in current to voltage conversion, affecting precision measurements.

**4.1.2. Bandwidth Extension of Classical Transimpedance Amplifier with High Pass Filter**

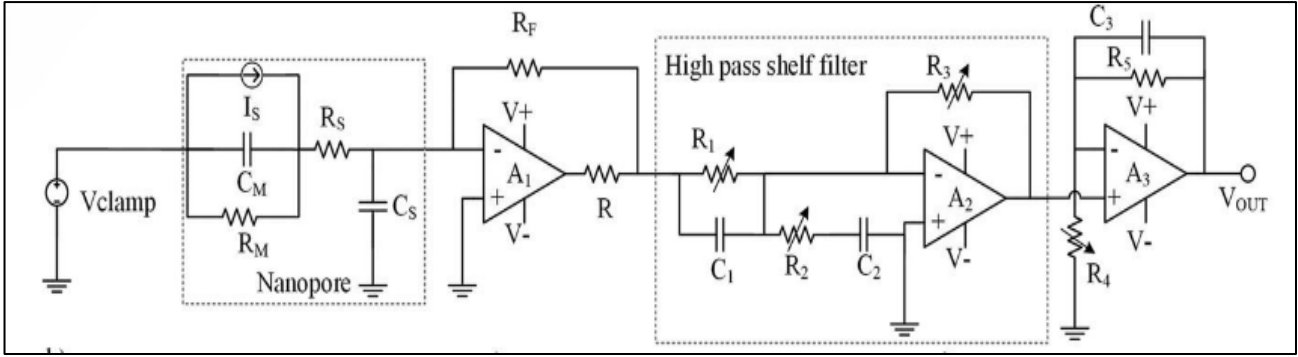

*Figure 6: Dual Stage Schematic of the Two-Stage, Low Noise Transimpedance Amplifier Architecture with Integrated High-Pass Filter.*

**4.1.2.1 Design and Implementation**
The amplifying circuit has been improved to achieve low noise current amplification using only the first stage amplifying circuit. To overcome the bandwidth limitation of the first stage Transimpedance Amplifier (TIA), several techniques such as cascade amplification, T-networks, capacitance compensation, and high pass shelf filters were considered. Among these, the inverting high pass shelf filter was chosen due to its effectiveness in bandwidth expansion and ease of frequency adjustment [32]. The circuit consists of adjustable components: $R_1$ and $R_2$ (potentiometers) along with capacitor $C_1$ for bandwidth tuning, while $R_3$ allows independent DC amplitude adjustment. A series resistor was integrated to enhance stability by minimizing oscillations and improving open loop output resistance. The high pass shelf filter is followed by an adjustable amplifier and a Bessel filter circuit, ensuring compatibility with the A/D data acquisition system's voltage input range while improving detection accuracy. To minimize external noise interference, the entire circuit is enclosed in an aluminum Faraday cage, significantly reducing noise levels and enhancing its suitability for nanopore detection.

**4.1.2.2 Performance**
The first stage TIA, built using the ADA4530-1 chip with a 1GΩ resistor, achieves a bandwidth corner frequency of 541 Hz. The high pass shelf filter extends the frequency compensation range from 466 Hz to 52 kHz, significantly broadening the current acquisition bandwidth. The final implementation supports a sampling frequency exceeding 50 kHz, meeting the detection requirements of nanopore sensors. Precise frequency response is achieved by carefully matching pole and zero locations using low tolerance resistors and capacitors, preventing frequency shifts and ensuring circuit stability. Noise power spectrum analysis demonstrates that the circuit delivers performance comparable to the industry standard Axon 200B, validating its effectiveness for high sensitivity applications.

**4.1.2.3 Challenges**
Designing subsequent amplification and filtering stages (adjustable amplifier and Bessel filter) to appropriately match the circuit output to the input voltage range of the analog to digital data acquisition system, ensuring optimized detection performance. Stability issues arising from oscillations necessitated the integration of a series resistor, which helped in improving the open loop output resistance and stabilize the system. Additionally, achieving precise frequency response required meticulous selection of circuit components to ensure accuracy while maintaining low noise. The challenge of designing a system that balances bandwidth expansion with noise reduction required an optimized circuit layout and careful tuning of compensation networks. Ensuring sufficient bandwidth (around 50 kHz) for detecting nanopore events was challenging because pulse signals above 50 kHz began degrading, transitioning from clear rectangular waves to triangular shapes, indicating bandwidth limitations.

### 4.1.3. Capacitive Feedback Integrated Current Acquisition circuit for Nanopore Based DNA Sequencing

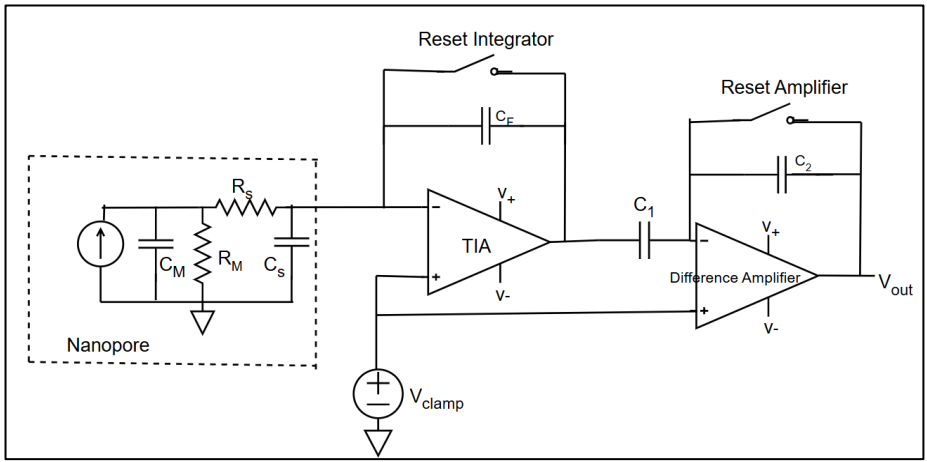

*Figure 7: Current Acquisition Circuit Featuring Capacitive Feedback Integration.*

#### 4.1.3.1 Design and Implementation

The amplifying circuit, as shown in Fig. 6, represents a system block diagram of a capacitive feedback integrated current acquisition circuit [32]. This design consists of a current integrator in the first stage, followed by an inverting post-amplifier stage to achieve enhanced signal amplification. The first stage operational trans conductance amplifier (OTA) is responsible for performing voltage to ground conversion, where the input current is integrated over time to generate an output voltage. Mathematically, the output voltage is given by:

$V_{out} = \frac{1}{C_F} \int_0^T i(t) dt,$

Where T represents the integration duration, i(t) is the measured input current, and $C_F$ serves as the integration capacitor. The bandwidth of the system is determined by the RC network formed by $C_F$ and $R_{in}$ which corresponds to the nanopore resistance. Given that nanopore resistance typically exceeds 1 GΩ, a capacitor in the range of 100 fF to 1 pF is recommended for optimal integration performance. To further enhance signal amplification, an additional post amplifier stage is introduced, where gain tuning relies on the ratio 'n', defined as $C_1 = n * C_2$. Precise tuning of this ratio is essential to maintain system stability and prevent oscillations.

#### 4.1.3.2  Performance

The integration-based detection mechanism ensures improved signal clarity by reducing noise, though it requires balancing temporal resolution. Longer integration times reduce noise and enhance signal accuracy but may limit the ability to detect fast changing signals. The measured nanopore current is calculated using:

$I_{nanopore} = \frac{C_F}{n} \frac{V_f - V_i}{\Delta t}$, where $V_f$ and $V_i$ correspond to the voltage values before and after the integration period, and $\Delta t$ represents the sampling interval. These reset switches play a crucial role in eliminating residual charges, ensuring measurement accuracy, and enhancing the stability of nanopore current detection. The Reset Integrator switch discharges the integration capacitor before each measurement cycle, preventing charge accumulation and establishing a consistent initial state. Similarly, the Reset Post switch resets the post amplifier stage, allowing proper execution of correlated double sampling (CDS) and improving measurement precision. To ensure that only the desired signal is processed within the detection bandwidth, a low pass filter is incorporated after the circuit, effectively removing high frequency noise. This filtering technique improves the signal to noise ratio (SNR) and enhances measurement reliability. Given its wide bandwidth characteristics, this architecture is particularly well suited for detecting weak photocurrents in fast changing signals, making it highly applicable in ASIC based designs where precise current detection is required.

#### 4.1.3.3 Challenges

Selecting capacitors in the femtofarad (fF) to picofarad (pF) range is challenging since tiny capacitance values are sensitive to parasitic effects and manufacturing tolerances, significantly impacting the accuracy, stability, and reproducibility of the measured current. Implementing the correlated double sampling method taking measurements immediately after reset and at the end of integration demands precise timing and low noise circuits. Any noise, timing inaccuracies, or drift between the two sampling points ($V_i$ and $V_f$) directly impacts current measurement accuracy. Achieving precise capacitor matching for predictable gain is challenging due to manufacturing variations and parasitic effects. The challenge lies in designing a control system capable of managing these operations efficiently while maintaining precise timing to avoid measurement errors.

### 4.1.4. A Low Noise Capacitive Feedback Amplifier with Bandwidth Extension for Nanopore based Analysis

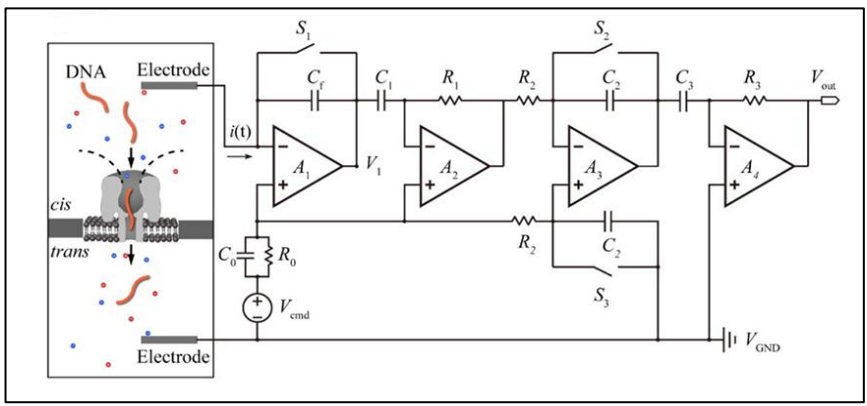

*Figure 8: Schematic Representation of a Low-Noise Amplifier for Nanopore Single Molecule Analysis.*

#### 4.1.4.1 Design and Implementation

The capacitor feedback low noise amplifier system has been developed to enhance current measurement accuracy in nanopore based experiments by minimizing thermal noise, which is a common drawback in traditional resistor feedback amplifiers. By employing a capacitor feedback configuration, the system achieves low noise amplification while integrating a differential integrator to suppress common mode noise caused by reference voltage fluctuations. This improves signal clarity and ensures accurate detection of ultra-low current signals, often below 10 pA. Additionally, a drift compensation mechanism has been integrated to prevent long term signal distortion and maintain measurement stability over extended recordings. The amplifier operates as a voltage clamp, maintaining a stable reference voltage while integrating the ionic current. Its first stage integrates the input current, producing an output voltage $V_1$ that is directly related to the charge accumulated in the feedback capacitor $C_f$, given by:

$$V_1 = \frac{1}{C_f} \int i(t)dt$$

To further refine the signal, a differential amplifier $A_2$ acts as a subtractor, removing any drift from the reference voltage. The output voltage $V_2$ which is a function of the reference voltage and integrated current, is expressed as:

$$V_2 = V_{ref} - \frac{R_1 C_1}{C_f} \int_0^t i(t)dt$$

A differential integrator $A_3$ then processes the difference between $V_2$ and the reference voltage, ensuring that the final output remains linear to the input current:

$$V_{out} = \frac{R_3 C_3 R_1 C_1}{C_f R_2 C_2} i(t)$$

To prevent saturation in the feedback capacitors ($C_f$ and $C_2$) CMOS switches $S_1$, $S_2$, and $S_3$ with low leakage currents have been implemented to discharge the capacitors without introducing additional noise. The processed signals, $V_1$ and $V_{out}$, are then digitized using the AD7626 ADC (Analog Devices), ensuring high-precision digital conversion of the analog measurements.

### 4.1.4.2 Performance

This amplifier system has achieved high sensitivity and a broad dynamic range, enabling precise detection of low current blockages and multi-level structures in single molecule events. The capacitor feedback design has effectively reduced thermal noise, while the differential integrator has eliminated common mode noise, significantly improving the signal to noise ratio (SNR). The amplifier maintains signal integrity across a wide bandwidth, even when using a small feedback capacitor (0.1 pF). Furthermore, high frequency noise suppression has been implemented to ensure clear signal acquisition, making the system well suited for nanopore based sensing applications. Detected and clearly differentiated between molecular translocation events and low amplitude bumping interactions, resolving two distinct Gaussian peaks (0.21 I/Io and 0.55 I/Io), which conventional systems failed to achieve. The combination of voltage clamping, differential integration, and low noise amplification allows for highly accurate current measurements, making this system an essential tool for nanopore based experiments requiring ultra-low current sensitivity.

### 4.1.4.3 Challenges

Selecting an extremely small capacitor (0.1 pF) to achieve high bandwidth is challenging due to sensitivity to parasitic capacitances and component variability. The design has required precise drift compensation to mitigate long term signal drift and prevent measurement inaccuracies. Distinguishing low amplitude blockage signals from the baseline noise presented a challenge, necessitating significant noise reduction to enable lower detection thresholds. Managing and suppressing high frequency and thermal noise required careful design, including a capacitor feedback structure, differential integration, and drift compensation. Preventing saturation of feedback capacitors ($C_f$ and $C_2$) was challenging, necessitating precise and low leakage CMOS switches (S1, S2, S3) for effective capacitor discharge without introducing additional noise.

### 4.1.5. A Hybrid Semi-Digital Transimpedance Amplifier with Noise Cancellation Technique for Nanopore Based DNA Sequencing

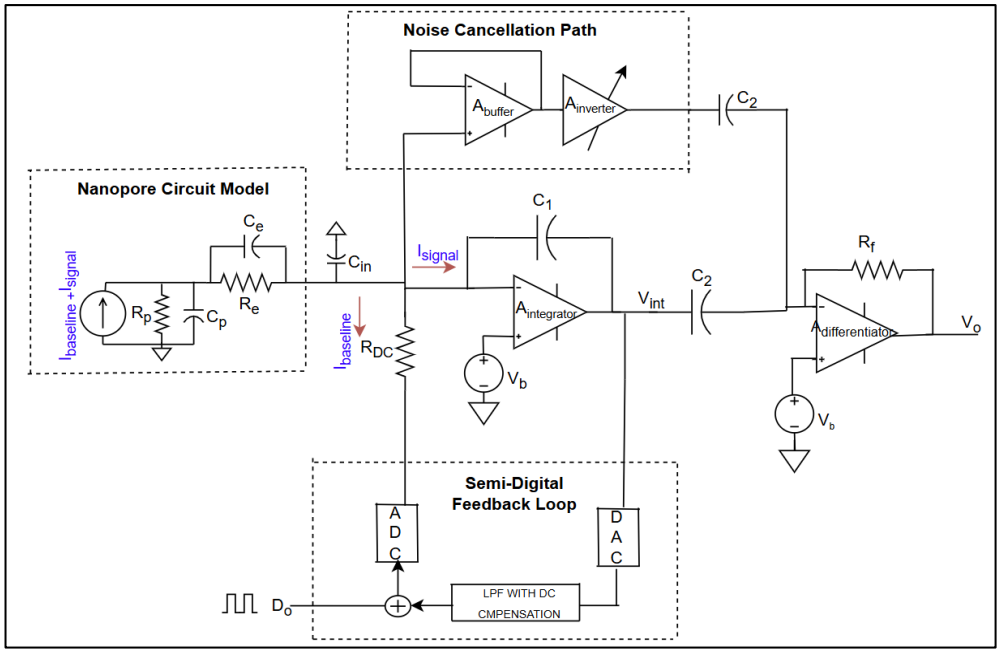

*Figure 9: Circuit Architecture of a High-Speed TIA Implementing Noise Reduction Mechanisms.*

#### 4.1.5.1 Design and Implementation

Hsu et al. introduced a Hybrid Semi Digital (HSD) Feedback System for a Transimpedance Amplifier (TIA), specifically designed for low noise, fast settling time, and continuous operation in nanopore based DNA sequencing [37]. This system incorporates an integrator differentiator design with semi digital feedback and feed forward noise cancellation to enhance both stability and noise performance. The signal path consists of a capacitive feedback integrator followed by a differentiator, ensuring a high flat gain bandwidth by canceling the pole with a zero. The mid band gain of the system is determined by the ratio of feedback capacitors and the differentiator resistance. This loop consists of an ADC, a digital low pass filter (LPF), and a DAC, dynamically adjusting the low cutoff frequency to accommodate nanopores with slow DNA translocation speeds. By eliminating bulky resistors and capacitors that introduce large variations, the system ensures precise frequency control without compromising stability. The noise characteristics of the HSD TIA are primarily influenced by input capacitance and voltage noise. The input capacitance, dominated by nanopore capacitance, OPAMP input capacitance, and parasitic cable capacitance, typically remains around 10 pF. Since reducing this capacitance is inherently constrained, the system implements a feed forward noise cancellation mechanism. A voltage sensing amplifier and an inverting amplifier with matched gain are employed to suppress voltage noise without distorting the input signal.

#### 4.1.5.2 Performance

The presented TIA successfully achieves a high flat gain bandwidth exceeding 950 kHz, while effectively maintaining a wide dynamic range capable of accurately handling input currents up to 10 pA. By balancing the noise and signal paths effectively, this method significantly lowers input referred noise compared to conventional TIAs. A key feature of the design is the semi digital feedback loop, which actively suppresses low frequency noise, including baseline current fluctuations and flicker noise, while maintaining phase margin stability above 45°. The settling behavior of the amplifier depends on the feedback loop bandwidth and low cutoff frequency. In response to a step input current, baseline current accumulation on the integrator capacitor can lead to saturation due to the slow discharge rate of the feedback loop. To mitigate this issue, an adaptive DC compensation current is introduced. A digital circuit in the FPGA detects step input occurrences and applies a compensation current of opposite polarity, preventing integrator saturation while maintaining continuous operation. This technique minimizes settling time by ensuring that the feedback and baseline currents remain balanced, eliminating the need for the low pass filter to stabilize before processing signals.

#### 4.1.5.3 Challenges

Designing the feed forward noise cancellation circuit necessitated exact matching of voltage signals to effectively suppress dominant voltage noise at high frequencies is challenging. Implementing adaptive DC compensation was essential to achieving rapid settling time, preventing integrator saturation, and maintaining continuous operation without the need for a reset network. Another challenge was ensuring that the amplifier responded quickly enough to capture transient DNA signals, which required sophisticated digital filters and real time compensation mechanisms. Additionally, input referred current noise, largely influenced by unavoidable parasitic capacitances, posed difficulties in maintaining low noise performance while sustaining high bandwidth. The digital feedback loop enabled continuous operation by precisely controlling the low frequency response, while the adaptive current compensation mechanism significantly improved settling time. These advancements make the HSD-TIA an optimal solution for nanopore based DNA sequencing and other high speed, high precision signal processing applications.

### 4.1.6. Low Noise Dual Channel Current Amplifier for DNA Sensing

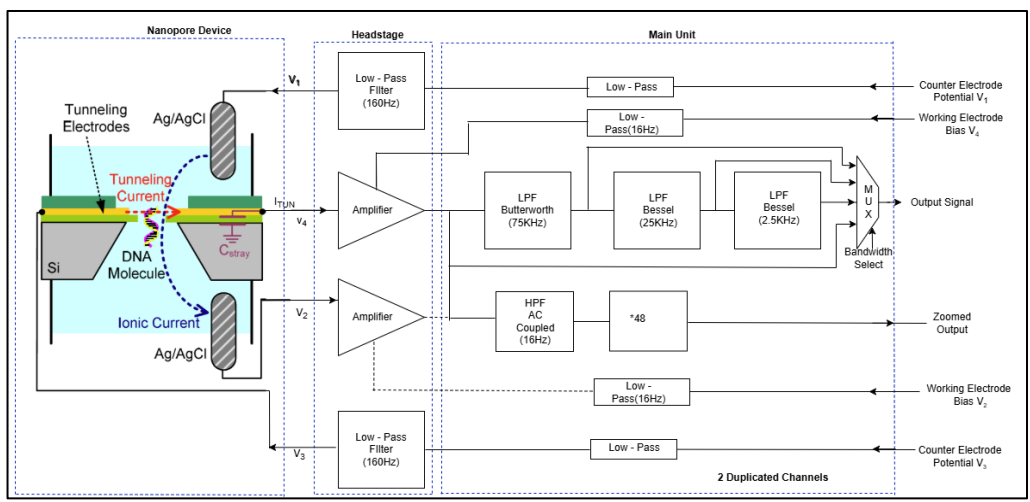

*Figure 10: Architecture of the nanopore characterization system featuring dual current-sensing amplifiers with configurable filters and integrated biasing circuits.*

**4.1.6.1 Design and Implementation**

Carminati et al. propose a dual channel low noise current amplifier designed for nanopore based DNA sensing [38]. The transimpedance amplifiers are enclosed in a compact metallic box (15 cm × 15 cm × 4 cm) and placed in a well-grounded Faraday cage to minimize noise, while the main unit houses power regulation and signal filters. The system derives ±12V and ±5V from a main ±15V dual power supply, with extensive filtering. Electrode bias potentials are independently adjustable, with on board jumpers allowing selection between grounding or external signal connections. Noise reduction is achieved through filtered biasing at 160 Hz for excitation electrodes ($V_1$, $V_3$)  and 16 Hz for reading electrodes ($V_2$, $V_4$), with adjustable bandwidth for dynamic waveforms. The amplifier outputs are low pass filtered through a cascade of 3rd order filters, including a 75 kHz Butterworth filter, a 25 kHz Bessel filter, and a 2.5 kHz Bessel filter, with a digitally controlled analog multiplexer for bandwidth selection. Additionally, an AC coupled zoomed output (16 Hz) provides ×48 amplification, improving resolution during the blockade phase, with a fast reset network preventing saturation. The current sensing system consists of two identical channels using a differential transimpedance amplifier with matched JFETs to achieve low voltage noise 1.5nV/√Hz while handling high input capacitance (~70 pF). The differential stage is biased with an 8 mA trimmable current and cascoded to a 300Ω resistive load for offset cancellation. A commercial operational amplifier (OP1) provides a total voltage gain of ~120 dB with a gain-bandwidth product of 1.7 GHz, ensuring a closed-loop bandwidth exceeding 75 kHz for input capacitance up to 22 nF.

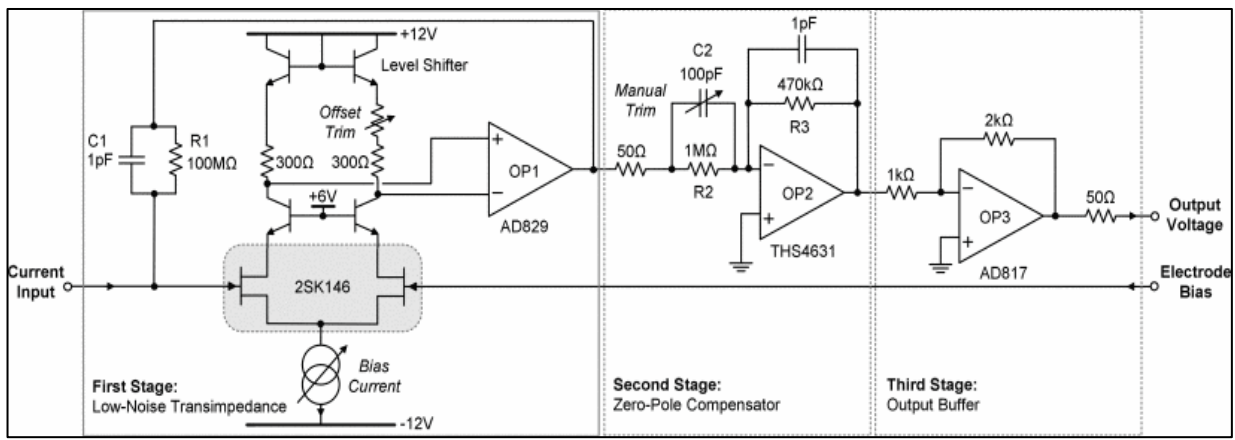

*Figure 11: Schematic diagram of a low noise transimpedance amplifier featuring integrator and differentiator stages, implemented with discrete JFET pairs. Reproduced with permission from [38]. Copyright (ICECS 2012).*

**4.1.6.2  Performance**

Experimental results confirm a flat frequency response, with a measured bandwidth of approximately 300 kHz, consistent with PSpice simulations. The gain for the ionic channel is 90 MΩ , while for the tunneling channel it is 92 MΩ, within expected tolerances. Noise analysis reveals distinct performance differences among the evaluated systems. Standard transimpedance amplifiers exhibit significantly higher current noise levels of approximately 400 fA/√Hz , limiting their utility in nanopore based sensing applications. In contrast, the Axopatch 200B amplifier, utilizing capacitive feedback, achieves an exceptionally low current noise of 0.7fA/√Hz, but requires frequent capacitive feedback resets, complicating continuous operation. The proposed amplifier system provides a favorable compromise, maintaining a low current noise level around 12fA/√Hz approximately 33 times lower than standard transimpedance amplifier and offering a voltage noise of 1 nV/√Hz, comparable to the Axopatch system. Although external capacitance slightly increases high frequency noise, the proposed design remains advantageous due to its high gain, low noise, extended bandwidth, and effective filtering. Standard transimpedance amplifiers face a resolution bandwidth trade off due to the feedback resistor ($R_1$) influencing gain, input noise, and bandwidth. To overcome this, a pole zero compensation method is implemented by adding a capacitor ($C_1$) in parallel with $R_1$, with $C_2$ set as $C_2 = C_1 * \frac{R_1}{R_2}$. A capacitive trimmer compensates for component tolerances, and the overall gain remains frequency independent, determined by $R_3 * \frac{C_2}{C_1}$, achieving approximately 94 MΩ. A third stage (OP3) provides additional gain (×2), and the system drives a shielded cable to the main unit, ensuring stability and noise immunity.

### 4.1.6.3 Challenges

Major design challenges included the need for meticulous manual tuning and bias trimming to ensure stable and optimal amplifier performance. The pole zero compensation network required careful adjustments to maintain consistent frequency response and gain stability. The primary challenges faced in designing this TIA included managing the high input capacitances (~550 pF to several nF) from nanopore electrodes, which significantly constrained achieving both high bandwidth and low noise simultaneously. Additionally, there was a fundamental resolution bandwidth trade off due to the use of a single feedback resistor, which inherently forced compromises between amplifier gain, noise performance, and achievable bandwidth. Variability in actual measured gains highlighted challenges related to the tolerances of feedback components.

### 4.1.7. Advanced Circuit Design of Wide Bandwidth Transimpedance Amplifier for extremely High Sensitivity Continuous Measurements

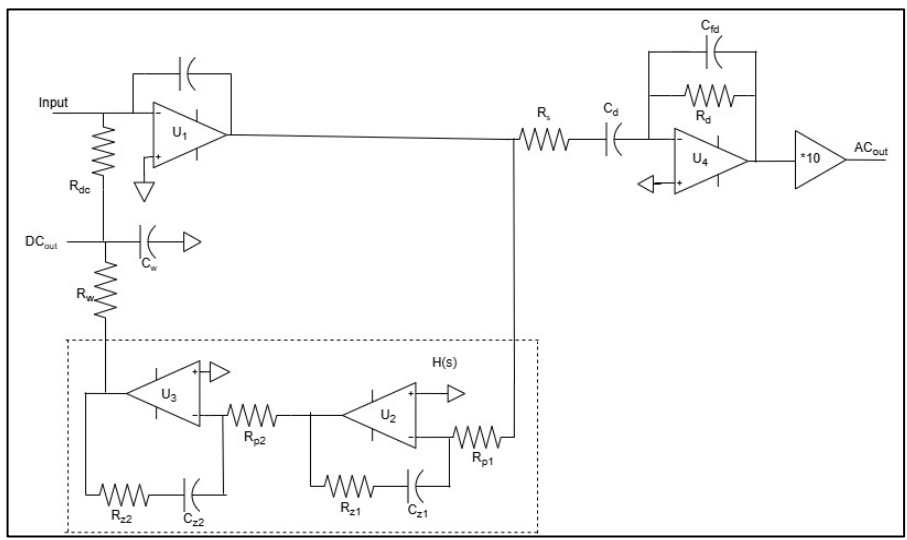

*Figure 12: Realization of a wide bandwidth, low noise transimpedance amplifier for ultra-sensitive continuous measurements using standard discrete components.*

**4.1.7.1 Design and Implementation**

On-Chip Implementation of Wide Bandwidth Transimpedance Amplifier for High Sensitivity Continuous Measurements, where the low frequency components, including baseline current and flicker noise, are discharged by the analog block, whereas in the hybrid digital semi approach discussed in 4.1.5, they are discharged by the digital block. The proposed transimpedance amplifier (TIA) incorporates a novel feedback network comprising amplifier H(s) and resistor $R_{dc}$ to discharge the input stationery current [39]. The function of H(s) is to ensure that DC variations at the integrator output (Node A) are transferred to Node B, facilitating the discharge of $C_i$ or the absorption of the device under test (DUT) standing current while simultaneously suppressing high frequency voltage changes to prevent signal interference at input. For stable operation, the cutoff frequency $f_m$ is set very low (a few Hz or lower), and H(s) provides at least a 90° phase shift at $f_m$, ensuring stability with $|G_{loop}(f_m)| = 1.15$. To maintain stability, H(s) is designed with matched poles and zeros, while additional high frequency poles are introduced with minimal impact on system performance. The amplifier's operational bandwidth extends from $f_m$ to $f_\gamma$, where for $f < f_m$, $R_{dc}$ efficiently carries the input current, preventing charge accumulation in $C_i$, and for $f > f_m$, the amplifier takes over. The upper frequency limit $f_t$ is determined by the operational amplifier gain bandwidth product (GBP), while $f\gamma$ is influenced by $R_{dc}$, $C_i$ and amplifier characteristics. The circuit ensures a low frequency response with a transfer function equivalent to $R_{dc}$ up to $f_m$, enabling effective DUT bias monitoring while decoupling the signal range from leakage current. The maximum AC input current depends on frequency and integrator saturation voltage, whereas the maximum DC input current is constrained by:

$$I_{DC,max} = \frac{V_{sat,H}}{R_{dc}}$$

A high $R_{dc}$ value helps reduce noise without significantly impacting bandwidth and does not require high precision, as H(s) adjusts Node B voltage. CMOS integration is feasible by replacing $R_{dc}$ with active components such as diodes or MOSFETs, making the design compatible with small capacitances and ultra-low bias currents, ideal for compact, low-noise CMOS implementations.

**4.1.7.2 Performance**

The noise analysis of the TIA considers contributions from operational amplifiers, $R_{dc}$, and H(s) When H(s) and $R_{dc}$ are neglected, noise input is primarily influenced by thermal noise, amplifier noise, and stray capacitance $C_p$ at the integrator input. The current gain $\frac{C_p}{C_i}$ effectively reduces differentiator noise, causing $R_{dc}$ thermal noise to appear as a higher resistance. Lowering $C_i$ reduces noise but limits the maximum signal frequency, while increasing $R_{dc}$ further decreases noise but restricts the measurable DC current, requiring a tradeoff between sensitivity and current range. To prevent high frequency noise amplification, a low pass filter $R_w$-$C_w$ is employed, ensuring stable operation. Stray capacitance $C_p$ significantly impacts noise performance, making short input cables and close placement of the amplifier to the DUT essential for noise reduction. By carefully balancing these parameters, the system achieves low noise across a broad bandwidth while effectively suppressing interference.

**4.1.7.3 Challenges**

Several challenges were encountered in the design of the transimpedance amplifier. Stability concerns arose due to the integrator stage introducing a pole at zero frequency, necessitating precise phase margin and gain control through H(s). Noise minimization was a critical factor, as thermal noise from $R_{dc}$, operational amplifiers, and stray capacitance $C_p$ had to be effectively managed. Controlling the bandwidth was also challenging, as it was necessary to ensure a stable low-frequency response while maintaining a flat high frequency performance. Managing leakage currents was essential to keep $V_{out}$ near zero for maximizing signal range and maintaining linearity. The selection of $R_{dc}$ enquired careful consideration to balance noise reduction with measurable DC current limitations. Additionally, integrating the system into a CMOS process required low power and compact design solutions, leading to the replacement of $R_{dc}$ with active components such as diodes or MOSFETs. Despite these challenges, the TIA achieves high stability, low noise, and wide bandwidth control. The feedback network ensures accurate signal amplification, and the inclusion of H(s) effectively reduces thermal and amplifier noise. The system efficiently manages leakage currents, keeping $V_{out}$ stable and enhancing the dynamic range. Furthermore, the design is well suited for CMOS applications, supporting small capacitances and ultra-low bias currents, making it ideal for compact, low-noise implementations. The flexible feedback network allows for adjustments without requiring precision components, simplifying implementation. Overall, the TIA achieves a stable, low noise performance with broad operational bandwidth, making it a robust solution for transimpedance applications.

### 4.1.8. Axon Axopatch 200B

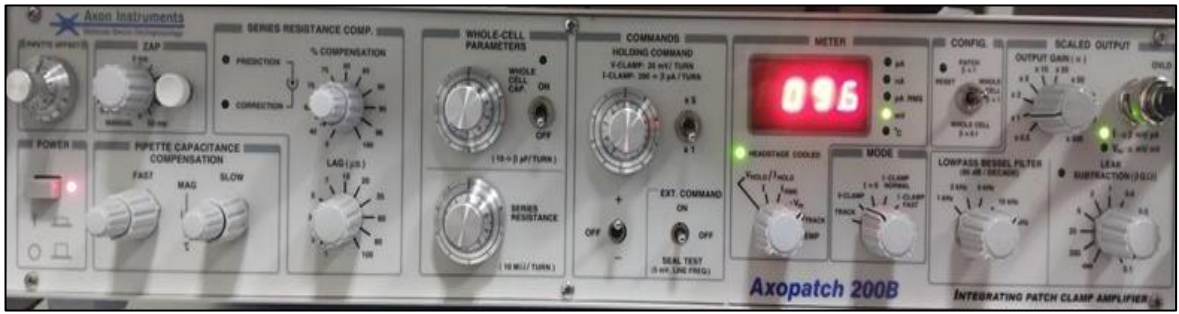

*Figure 13: Axopatch 200B off shelf TIA*

**4.1.8.1 Design and Implementation**
The Axopatch 200B Amplifier employs innovative capacitor feedback technology to minimize noise, making it highly suitable for single channel and whole cell patch clamp recordings. It integrates active cooling of the critical electronic components down to approximately -15°C to significantly reduce thermal noise. The amplifier includes built-in capacitance compensation for patch, whole cell, and loose patch modes, ensuring versatility across various recording setups. A slim head stage design facilitates better electrode access and ease of positioning. Additionally, advanced features such as real time leak subtraction, dual speed current clamp modes, and specialized series resistance compensation techniques further enhance its adaptability and operational convenience. Integrated functionalities such as a built in ZAP feature allow controlled rupture of cell attached patches, adding convenience during whole cell experiments.

**4.1.8.2 Performance**
The Axopatch 200B demonstrates exceptionally low open-circuit noise (as low as 0.13 pA RMS at 10 kHz), superior to conventional resistive feedback amplifiers. The capacitor feedback configuration offers improved dynamic range and reduced frequency of system resets, allowing precise measurement of sub picoampere currents. It supports a broad bandwidth range (up to 100 kHz), providing accurate and high-fidelity data capture and exhibits excellent stability under substantial input capacitance loads (up to 1000 pF), enhancing reliability in bilayer and complex cell recording setups. This performance is complemented by enhanced seal testing capabilities, advanced pipette offset corrections, and efficient cell capacitance compensation for recording stability and accuracy. Features versatile external command inputs, enabling simultaneous application of multiple experimental stimuli, crucial for advanced electrophysiological studies.

### 4.1.8.3 Challenges

Despite its advancements, the Axopatch 200B encounters certain operational challenges. The integration-based capacitor feedback approach necessitates periodic resets, though optimized to be brief (approximately 50 µs) and infrequent. The handling of large currents, especially during substantial voltage steps in bilayer experiments, requires careful management to maintain stability and accuracy. Additionally, maintaining precise temperature control for active cooling to minimize noise involves added complexity in practical laboratory setups. During large command voltage steps, managing transient current delivery requires careful handling to prevent instability or electrode damage.

After analyzing various off-chip TIA architectures, their performance metrics, and design challenges, it is crucial to compare them based on parameters such as gain, bandwidth, noise performance, power consumption, and channels for specific applications. Table II provides a comparative summary of various off-chip TIA methods, highlighting their advantages and limitations.

| Parameter | 4.1.1. | 4.1.2. | 4.1.3. | 4.1.4. | 4.1.5. | 4.1.6. | 4.1.7. | 4.1.8. |
|---|---|---|---|---|---|---|---|---|
| Parasitic capacitance | 50pF | 45pF | 50pF | 50pF | 10pF | 70pF | 8pF | 1000pF |
| Supply Voltage | ±5 V | 3-5V | ±2 V | 1.8 V | ±5 V | ±15 V | ±10V | ±10V |
| Gain | 180dBΩ | 180dBΩ | 180dBΩ | 200dBΩ | 150dBΩ | 160dBΩ | 146dBΩ | 0.5x-500x |
| Input Referred Noise | —— | $28.2\frac{fA}{\sqrt{Hz}}$ | $10\frac{fA}{\sqrt{Hz}}$ | $10\frac{fA}{\sqrt{Hz}}$ | $8.5\frac{fA}{\sqrt{Hz}}$ | $12.12\frac{fA}{\sqrt{Hz}}$ | $54.7\frac{nA}{\sqrt{Hz}}$ | $1.45\frac{fA}{\sqrt{Hz}}$ |
| Power Consumption | —— | —— | —— | —— | 65 mW | —— | 640mW | —— |
| Bandwidth | 530Hz | 52 kHz | 50KHz | 3 KHz | 950KHz | 75 KHz | 1.4MHz | 100KHz |
| Current Range | 1nA | 1nA | ±1.5nA | ±10 nA | 10pA-55 pA | ±100nA | ±10nA | 200nA |
| channels | 1 | 1 | 1 | 1 | 1 | 2 | 1 | 1 |

*Table II: Performance summary and comparison of previous works off chip approaches*

## 4.2. On-Chip Transimpedance Amplifier for Nanopore Based Systems

### 4.2.1. Design Methodology of Integrator - Differentiator Block

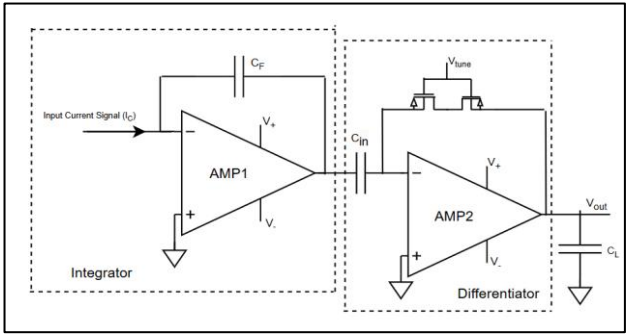

*Figure 14: Block diagram of proposed integrator-differentiator block.*

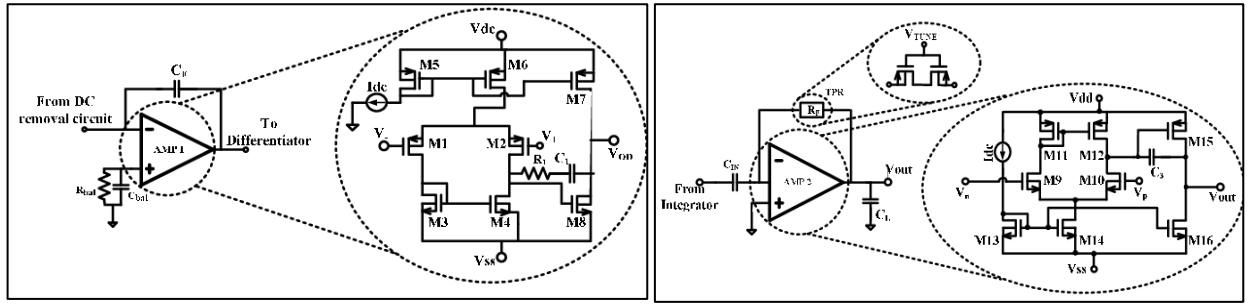

*Figure 15: Implementation of integrator and differentiator circuits using op-amps AMP1 and AMP2, respectively. Reproduced with permission from [40]. Copyright 2019 DevIC.*

**4.2.1.1 Design and Implementation**

The proposed integrator differentiator transimpedance amplifier (TIA) is designed for sensing small currents while addressing high noise and power consumption issues found in conventional architectures, such as those introduced by Sani et al [40]. The circuit is implemented in 0.18 µm CMOS technology, optimized for low power consumption and variable bandwidth. A MOS based pseudo resistor in weak inversion replaces the conventional off-chip resistor, enhancing the bandwidth noise trade off. Additionally, a Tunable Pseudo Resistor (TPR) in the differentiator's feedback path enables variable gain, improving adaptability in different sensing applications. TIA comprises two main stages: an integrator with capacitive feedback and a differentiator with resistive feedback, implemented using operational amplifiers $AMP_1$ and $AMP_2$, respectively. The TPR in the differentiator feedback loop provides tunable bandwidth control. The output voltage of the TIA is given by:

$$V_{OUT} = I_{IN} * R_F \left(1 + \frac{1}{j*2*\pi*R_F*C_F}\right)$$

where RF and CF define the gain and frequency response of the system. The overall voltage gain is expressed as $A_V = \frac{V_{OUT}}{I_{IN}} = R_F * \frac{C_{IN}}{C_F}$ and the -3dB bandwidth is determined by $BW_{-3db} = \frac{g_{mo}}{C_L}$ where $g_{mo}$ is the transconductance of the TIA and $C_L$ is the load capacitance. The integrator stage in Fig.13. consists of $AMP_1$, a feedback capacitor $C_F$, and a differential pair with current mirrors. The high W/L ratio in the input differential pair reduces flicker noise and enhances efficiency. The differentiator stages in Fig.14. implemented using $AMP_2$, incorporates an input coupling capacitor( $C_{IN}$ ) in order to obtain a flat gain, a TPR, and an op amp. The TPR, made of PMOS transistors, dynamically adjusts resistance through a control voltage $V_{tune}$, enabling tunability in gain and bandwidth.

#### 4.2.1.2 Performance

The proposed TIA effectively reduces noise while ensuring a broad operational bandwidth. The differentiator stage functions as a high pass filter, suppressing low frequency 1/f noise, which is a major limitation in traditional TIAs. By implementing a higher feedback resistance $R_F$, the overall noise contribution is lowered, while minimizing input capacitance $C_F + C_{IN}$ further enhances noise performance.

The noise power spectral density (PSD) of the system is given by $i_{IN}^2 = i_D^2 + i_N^2$ where $i_D^2$ represents the nanopore circuit noise (including flicker and thermal noise), and $i_N^2$ corresponds to the TIA interface noise. Since $i_D^2$ is negligible, the dominant noise contributor is $i_N^2$, which arises from the CMOS operational amplifier in the conventional TIA. To further suppress thermal and flicker noise, the first pole of the differentiator helps define the maximum bandwidth, eliminating the traditional noise bandwidth trade off. Compared to previous designs [39], the TPR based differentiator feedback offers better tunability and noise control, while active feedback replaces passive circuitry, reducing complexity and layout area.

#### 4.2.1.3 Challenges

The integrator differentiator TIA encountered several design challenges, primarily related to integrator saturation, noise optimization, and bandwidth stability. One of the most critical issues was integrator saturation, which occurs due to fast switching baseline currents and DC components, degrading frequency response. Managing feedback capacitor switching $C_F$ was necessary to prevent clock feedthrough, ensuring a stable output. Noise minimization required careful transistor sizing and circuit optimization, particularly in the differentiator stage, where the sub threshold operation of transistors helped reduce flicker noise. The challenge of balancing bandwidth and stability was addressed using a tunable pseudo resistor (TPR), which allowed fine adjustments in gain and frequency response without additional passive components. Another key challenge was DC mismatch between the integrator and differentiator, which could introduce offset errors and affect signal integrity. The input coupling capacitor $C_{IN}$ effectively eliminates these mismatches, ensuring consistent performance across a wide frequency range. Additionally ensuring stability between the integrator and differentiator stages was challenging due to DC mismatch, requiring the use of coupling capacitors and precise pole-zero placement.

### 4.2.2. A 90nm CMOS Transimpedance Amplifier Design for Nanopore based DNA Nucleotide Sequencer

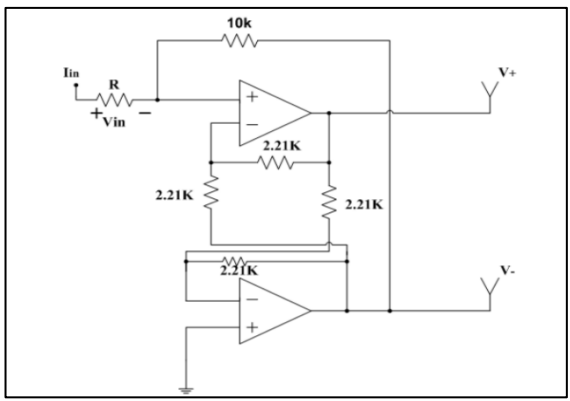

*Figure 16: Input current to differential voltage conversion circuit.*

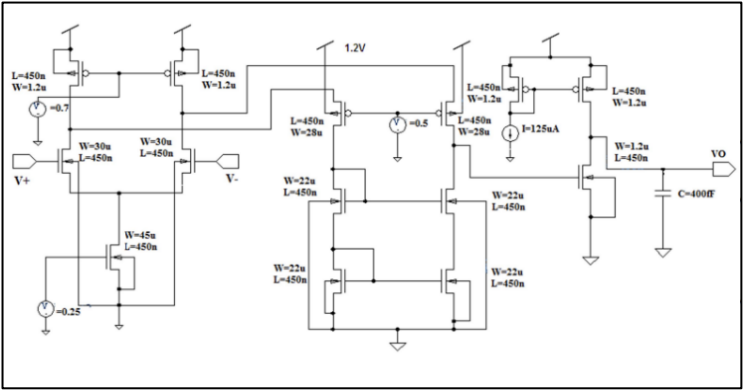

*Figure 17: Circuit diagram of the transimpedance amplifier for nanopore DNA sequencing, featuring a common-source amplifier in the second stage. Reproduced with permission from [41].Copyright (ICST), Kolkata, India, 2012*

**4.2.2.1 Design and Implementation**

The designed transimpedance amplifier (TIA) [41] achieves a high transimpedance gain of approximately 1 MΩ with a bandwidth of around 1 MHz, making it suitable for detecting low magnitude nanopore currents (~1 nA). To ensure low noise and a high signal to noise ratio (SNR), the TIA adopts multistage architecture. The nanopore current is first converted into a differential voltage as shown in Fig.15. using a single ended current to differential voltage converter. The first amplification stage is implemented using a differential folded cascode topology as shown in Fig.16. providing a gain of ~40 dB, which helps minimize noise contributions from subsequent stages. The second stage is a common source (CS) amplifier, contributing an additional ~20 dB gain. An optional third stage, employing a source follower (common drain) configuration, reduces output resistance, ensuring off-chip compatibility. The amplifier is designed using the 90 nm IBM CMOS process, operating at a 1.2V supply voltage with an input capacitance of 3 pF, making it suitable for a 30-nm nanopore membrane. The entire circuit was constructed and simulated using Tanner VLSI CAD tools (version 12). AC, transient, and noise analyses confirmed their effectiveness. The output successfully differentiates DNA nucleotides, demonstrating its feasibility for nanopore DNA sequencing applications.

**4.2.2.2 Performance**

In this design TIA efficiently converts picoampere level current variations into millivolt range voltage readings. The design achieves a total gain of 60 dB with a unity gain bandwidth of approximately 200 MHz. The wide unity gain bandwidth of 200 MHz further demonstrated its suitability for high speed nanopore applications, making it an effective solution for DNA sequencing. Designing for a nanopore with ~3 pF input capacitance required careful handling to maintain amplifier stability and bandwidth. Noise considerations were critical due to the low magnitude currents (~1 nA) being amplified. Simulation results confirmed that the TIA reliably distinguishes DNA nucleotides based on nanopore current variations, validating its effectiveness in high speed nanopore based biosensing applications.

**4.2.2.3 Challenges**

Designing the 90nm CMOS TIA involved several technical challenges. The primary challenge was amplifying extremely low nanopore currents while maintaining high gain and low noise. The noise factor is defined as $\frac{SNR_{in}}{SNR_{out}}$ where fluctuations in the source resistor Rn introduce current variations, making stable amplification challenging. The multistage design introduced additional implementation complexity, particularly in the first stage differential folded cascode topology and the second stage common source amplifier. Managing noise at each stage while maintaining bandwidth stability requires careful circuit optimization.

### 4.2.3. High precision low Power DNA Readout Interface Chip (RIC) for Multi-Channel Nanopore Applications

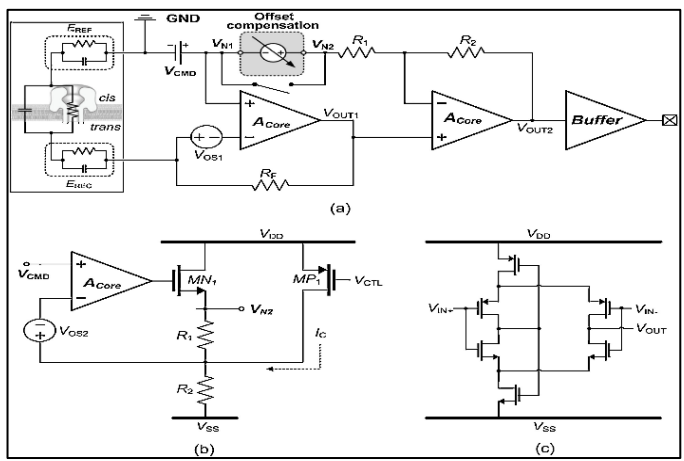

*Figure 18: Proposed DNA RIC Architecture and Core Amplifier Design}[42] (a) Proposed DNA RIC architecture with rf-TIA and high input impedance non inverting amplifier (b) offset cancellation block using $V_{CTL}$ to minimize offset, (c) core amplifier schematic with high gain bandwidth product, good phase margin, and low power dissipation at ±1.5 V. Reproduced with permission from reference [42]. Copyright Sensors and Actuators B: Chemical 2016.*

#### 4.2.3.1 Design and Implementation

A high precision and low power DNA readout interface chip (RIC) has been designed for multichannel nanopore applications. The proposed low offset, low power, and low noise DNA RIC [42] integrates an offset compensation block (OCB) and a non-inverting amplifier to enhance signal accuracy while minimizing unwanted deviations. The OCB compensates for DC and offset deviations, ensuring precise detection of ionic current variations caused by DNA translocation through a nanopore. When the switch is activated as shown in Fig.17. the input offset voltage $V_{OS1}$ and ionic current variation $I_N$ are amplified, potentially restricting the dynamic output range. To address this limitation, the OCB generates a compensation voltage $V_{OCB}$ that counteracts these deviations, thereby improving measurement accuracy. Additionally, variable current source $I_C$ further cancels residual offset effects, ensuring stable operation. A self-biased differential operational transconductance amplifier (OTA) is employed to achieve high gain and low noise using a current reuse technique, which enhances power efficiency. The OTA's NMOS and PMOS input pair transistors contribute to effective transconductance, providing a high open loop gain. The input referred noise, primarily attributed to thermal and flicker noise, is minimized by optimizing the aspect ratios of the input transistors. The design also reduces parasitic noise and interference by minimizing wiring complexity and power dissipation. A non-inverting amplifier is implemented in the second stage, providing high input impedance, which eliminates the need for an output buffer, thereby reducing power consumption and circuit complexity. The DNA RIC is fabricated using a 0.35 µm CMOS process, enabling low power and low noise operation. With a compact die area of 1.1 mm × 0.155 mm, the design is optimized for high precision nanopore based DNA sensing, demonstrating improved signal fidelity, power efficiency, and measurement stability.

#### 4.2.3.2 Performance

The proposed DNA RIC TIA effectively reduces offset voltage, enhances noise performance, and improves power efficiency while maintaining high transimpedance gain and measurement accuracy. The optimized self-biased OTA and large input transistors (PMOS & NMOS) contribute to input referred noise reduction, achieving a low noise level of 3.46pA RMS. The TIA provides a high transimpedance gain of 173.9 dBΩ within a 10 kHz bandwidth, allowing precise detection of DNA translocation blockade currents 50 –150 pA. Power efficiency is significantly enhanced by eliminating high current output buffers and employing a current reuse technique, making the system scalable for multichannel nanopore applications. Furthermore, an 8$^{th}$ order Bessel low pass filter (LPF) effectively suppresses high frequency noise, improving the signal to noise ratio (SNR) to 24.8 dB. The compact CMOS die (1.1 mm × 0.155 mm) ensures suitability for portable and miniaturized DNA sequencing applications, offering an optimal balance of high sensitivity, low power consumption, and scalability.

**4.2.3.3 Challenges**

The design of the DNA RIC TIA presented multiple challenges, particularly in offset voltage management, noise reduction, and power efficiency. The input offset voltage $V_{OS1}$ and DC deviation initially constrained the dynamic output range, necessitating the integration of an offset compensation block (OCB) to eliminate deviations while preserving signal integrity. Managing low frequency flicker noise and thermal noise at mid-range frequencies was essential for accurate detection of ionic currents. The integration of a self-biased OTA and large input transistors effectively suppressed input-referred noise, yet maintaining high gain while minimizing power consumption remained a challenge, especially for multichannel applications where multiple TIAs operate simultaneously. The nanopore and circuit layout introduced parasitic capacitance, which led to high frequency noise, requiring careful circuit optimization. Additionally, the ±5% tolerance of Nwell resistors in the 0.35 μm CMOS process introduced minor gain variations, necessitating calibration to ensure performance consistency.

**4.2.4. Nanopore DNA Sensors In CMOS With On-Chip Low Noise Preamplifiers**

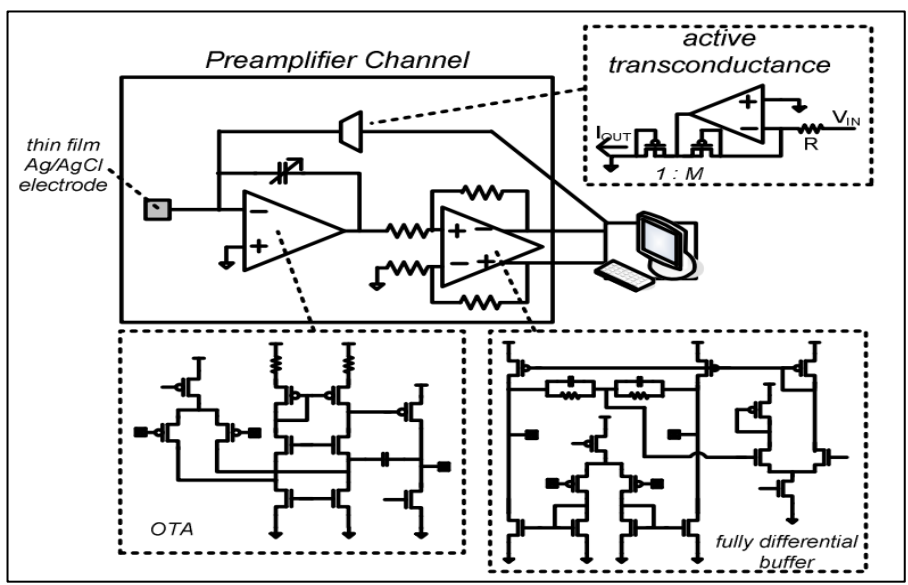

*Figure 19: Preamplifier circuit topology. Reproduced with permission from reference [43]. Copyright 2011 16th International Solid-State Sensors.*

#### 4.2.4.1 Design and Implementation

The custom preamplifier is a modified transimpedance amplifier (TIA) that incorporates a charge sensitive integrator with an active transconductance feedback loop instead of a traditional passive resistor. The input stage consists of a folded cascode operational transconductance amplifier (OTA) with a gain bandwidth product of 100 MHz. The feedback trans conductor is designed using matched PMOS transistors to ensure low noise current division, which improves stability and reduces signal distortion. A load compensated fully differential buffer is included to drive the output efficiently while preserving signal integrity. The system exhibits a flat low frequency response followed by a single pole, which is compensated by a cascaded filter with a zero, ensuring a flat frequency response before digitization. The amplifier operates across a wide frequency range from DC to beyond 1 MHz, with a pole zero pair at 25 kHz, achieving a gain of 106 MΩ while keeping the input capacitance below 2pF. The TIA is fabricated on a 3 mm × 3 mm die using a 0.13 µm 1.5V mixed-signal CMOS process. It integrates eight identical preamplifier channels, each occupying 0.3 mm² and consuming 5mW per channel. The integration of Ag/AgCl electrodes and solid state nanopores minimizes parasitic impedances, significantly reducing noise and improving measurement accuracy. This monolithic integration enhances the amplifier's effectiveness in high throughput nanopore based sensing applications.

#### 4.2.4.2 Performance

The noise floor of the TIA preamplifier is measured by dividing the output noise by the gain, with results aligned with simulation predictions. The input referred noise level is 12 fA/√Hz, equivalent to a 110 MΩ resistor, and is primarily attributed to the thermal noise in the feedback trans conductor, which remains independent of gain. The integration of Ag/AgCl electrodes and solid state nanopores effectively reduces parasitic impedances caused by external electronics, thereby minimizing high frequency noise. This monolithic integration offers a robust platform for high throughput single molecule sensing using nanopore sensor arrays. To further enhance performance, the folded cascode OTA with a 100 MHz gain bandwidth product maintains a stable gain response across a wide frequency range. Additionally, microfabrication optimizations successfully reduce membrane capacitance to less than 1 pF, improving sensitivity and noise performance.

#### 4.2.4.3 Challenges

The development of the transimpedance amplifier (TIA) for nanopore based DNA sensing presented several critical challenges. Amplifying weak, high impedance signals while minimizing noise from sources such as flicker noise, Johnson noise, and amplifier voltage noise proved complex. The system's limited bandwidth (10–50 kHz) and high membrane capacitance (≥ 300 pF) introduced stability concerns, requiring meticulous tuning of the charge sensitive integrator and transconductance feedback loop. Parasitic capacitance from interconnects further elevated the noise floor, complicating signal fidelity. Integration within the CMOS process posed additional constraints related to layout and component matching. Maintaining electrode stability, particularly for Ag/AgCl interfaces, was essential to prevent signal drift and ensure consistent performance. High power consumption was another limiting factor, driven largely by the need for an output buffer to support low resistance loads. On-chip area efficiency was reduced due to complex analog front end circuitry. Moreover, the amplification of input referred to offset voltages especially by the differential amplifier in the second stage negatively impacted the dynamic range and, in extreme cases, led to output saturation.

## 4.2.5. On-Chip Implementation of Wide Bandwidth Transimpedance Amplifier for Extremely High Sensitivity Continuous Measurements

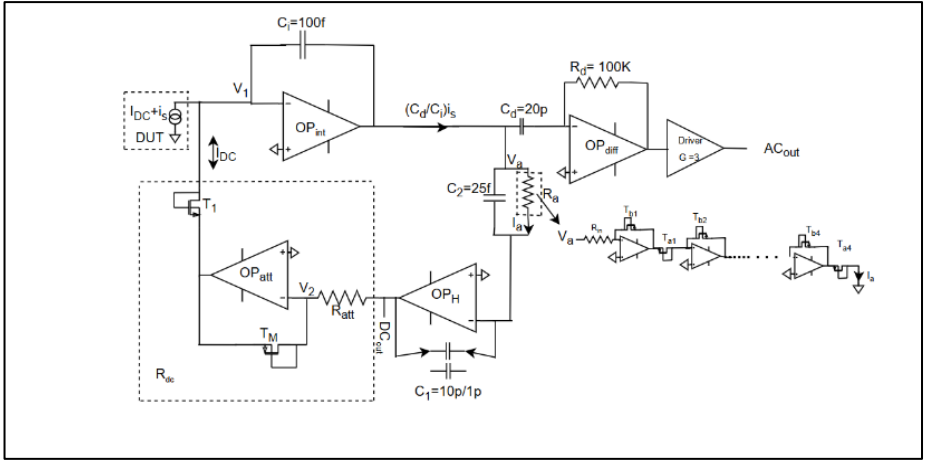

*Figure 20: Schematics of the transimpedance amplifier prototype with active network to drain the dc input current. The resistor $R_{dc}$ is implemented by the current reducer reported inside the dashed box on the lower left, while the large resistor $R_a$ is implemented by cascading 4 current reducer system.*

### 4.2.5.1 Design and Implementation

This is the on-chip implementation of the design discussed in 4.1.7, where the concept remains the same, but it is now realized on-chip. The proposed transimpedance amplifier (TIA) is implemented in CMOS 0.35 µm technology and is designed to detect weak currents in molecular and nanodevice applications. To achieve low noise, high input impedance, and wide dynamic range, the amplifier adopts an integrator differentiator architecture. Instead of relying on a traditional feedback resistor, a controlled capacitive feedback loop is used, which eliminates standing current and improves overall stability. This modification enables accurate measurement of currents as low as a few tens of nanoamperes while maintaining low noise levels. In biomolecular sensing applications, a key challenge is accurately measuring nanoampere level leakage currents. Conventional TIAs often suffer from noise, bandwidth limitations, and charge accumulation. To overcome these issues, the integrator stage is enhanced with a continuous discharge network that manages standing DC currents without degrading high-frequency response. This is achieved by incorporating an additional feedback loop containing a high-value active resistance $R_{dc}$, which supports low frequency signal processing while preserving high frequency performance. The active resistor $R_{dc}$ is implemented using a low noise linear trans conductor composed of matched MOSFETs $T_1$ and $T_M$, instead of a basic transistor-based approach. This configuration ensures consistent frequency behavior and loop stability. Both transistors are arranged in a source-well short circuit, where a negative gate-source voltage $V_{GS}$ forces them to operate as PMOS diodes. Their forward-biased drain-well junctions behave as p–n junction diodes, maintaining stable current density. By sizing $T_M$ to be M times larger than $T_1$, the circuit emulates a linear resistor with an equivalent resistance of 45 MΩ, delivering high linearity and low noise. Additionally, the use of active current reducers allows for an effective resistance in the hundreds of GΩ, further enhancing the amplifier's ability to sense ultra low currents. To ensure wide bandwidth and loop stability in the differentiator stage, an active Miller compensation technique is employed. Instead of using a conventional feedback capacitor, a deliberate zero is introduced to cancel the external pole introduced by $R_d$ and $C_d$. The inclusion of a small 1 pF Miller capacitor effectively reduces silicon area while maintaining the desired wideband response.

#### 4.2.5.2. Performance

An active low noise resistor was implemented using matched MOSFETs, providing a linear resistance of approximately 45 MΩ and an equivalent noise resistance of around 6.5 GΩ, making it suitable for ultra-low current operation. To reduce flicker noise, larger transistor areas were used, leveraging the fact that flicker noise decreases with increased device area. The design maintained stable feedback with a phase margin greater than 45° and ensured high linearity across the full input range. The noise performance of the high value resistor $R_{dc}$ was further optimized by minimizing thermal noise using a scaling factor $M^2$ applied to the physical resistor $R_{att}$. At low frequencies, shot noise originating from the MOSFET $T_1$ becomes dominant, while flicker noise ($1/f$ noise) is minimized by adopting modern MOSFET processes. The feedback network H(s) was carefully designed to include two poles and one zero. A low frequency pole enhances stability, while a high frequency zero boosts gain, improving both signal stability and bandwidth. This design ensures that the integrator output remains stable even in the presence of DC current variations. The forward amplifier in the integrator stage plays a crucial role in overall performance by influencing noise characteristics, high frequency bandwidth, and DC voltage stability at the input node. To reduce flicker noise, PMOS transistors were used in the input differential pair, due to their inherently lower flicker noise, high input impedance, and low gate bias currents. The input referred noise is influenced by capacitance at the inverting input, given as $e_{ni}^2 = (2\pi f)^2 (C_1 + C_{DUT} + C_{gate})^2 i_{ni}^2$, where $C_{gate}$ (transistor gate capacitance) and $C_{DUT}$ (device under test capacitance) must be minimized to reduce high frequency noise. The gate capacitance is further optimized as $C_{gate} = C_{ox} WL$ where W and L define the transistor dimensions. The differentiator stage addresses stability and bandwidth issues caused by high $R_d$ and $C_d$, which introduce a pole at 80 kHz. Instead of using a feedback capacitor for traditional compensation, a zero is introduced via a Miller compensation network, effectively canceling the external pole at $\frac{1}{2\pi R_d C_d}$. This method enhances stability and bandwidth, ensuring the amplifier maintains a wideband response with minimal noise.

#### 4.2.5.3 Challenges

Designing a low noise, high transimpedance amplifier involved addressing several key challenges to ensure optimal performance in terms of gain, bandwidth, and stability. One of the primary concerns was minimizing noise without compromising the amplifier's high gain requirements. Traditional feedback resistors, while commonly used, contribute to significant thermal noise, whereas MOS based pseudo-resistors, although better in some aspects, often limit bandwidth and can introduce instability. To effectively handle extremely low leakage currents in the femtoampere to nanoampere range, a continuous discharge mechanism was incorporated. This approach prevented charge accumulation at the input node, thereby preserving long-term stability and preventing output drift. Achieving wideband operation required careful pole zero compensation, which was implemented through a Miller feedback network, allowing precise control of the frequency response. The amplifier was integrated using CMOS 0.35 µm technology, which introduced additional design constraints, particularly in achieving high linearity and low power consumption while preserving a stable gain-bandwidth product. Within the integrator stage, the forward amplifier played a critical role by providing accurate DC voltage control and contributing minimal noise to the system. To further reduce current noise, an active resistor was employed, offering equivalent resistance in the GΩ range. This not only ensured low current noise at very small input currents but also minimized shot noise for currents above a few picoamperes, resulting in a high-performance, low-noise amplification system optimized for sensitive applications.

After discussing the key aspects of different on-chip TIA architectures, their performance metrics, and design challenges, it is essential to compare them based on critical parameters such as gain, bandwidth, noise performance, power consumption, and channels for specific applications. Table III provides a comparative summary of various off-chip TIA methods, highlighting their advantages and limitations.

| Parameter | 4.2.1. | 4.2.2. | 4.2.3. | 4.2.4. | 4.2.5. |
|---|---|---|---|---|---|
| **Technology** | 0.18 μm | 90nm | 0.35 μm | 0.13 μm | 0.35μm |
| **Parasitic capacitance** | 50nF | 3pF | ───── | 1pF | 0.5pF |
| **Supply** | 1.8V | 1.2V | ±1.5 V | 1.5V | ±1.5 V |
| **Gain** | 90.8 dBΩ | 60 dBΩ | 174 dBΩ | 160.56 dBΩ | 155.56 dBΩ |
| **Input Referred Noise** | $0.57 \frac{pA}{\sqrt{Hz}}$ | $20 \frac{fA}{\sqrt{Hz}}$ | $49.6 \frac{fA}{\sqrt{Hz}}$ | $12 \frac{fA}{\sqrt{Hz}}$ | $4 \frac{fA}{\sqrt{Hz}}$ |
| **Power Consumption** | 57.3μW | ───── | 213μW | 5 mW | 45 mW |
| **Bandwidth** | 100 KHz | 200KHz | 10KHz | 5MHz | 4MHz |
| **Current Range** | 10μA-2nA | ±1nA | 50pA - 150pA | 10 pA-55 pA | <10pA |
| **Channels** | 1 | 1 | 2 | 8 | 1 |
| **Area** | ───── | ───── | $0.1705 mm^2$ | $9 mm^2$ | $0.35 mm^2$ |

*Table III: Performance summary and comparison of previous works on chip approaches*

## 5. CHALLANGES AND FUTURE ADVANCEMENTS

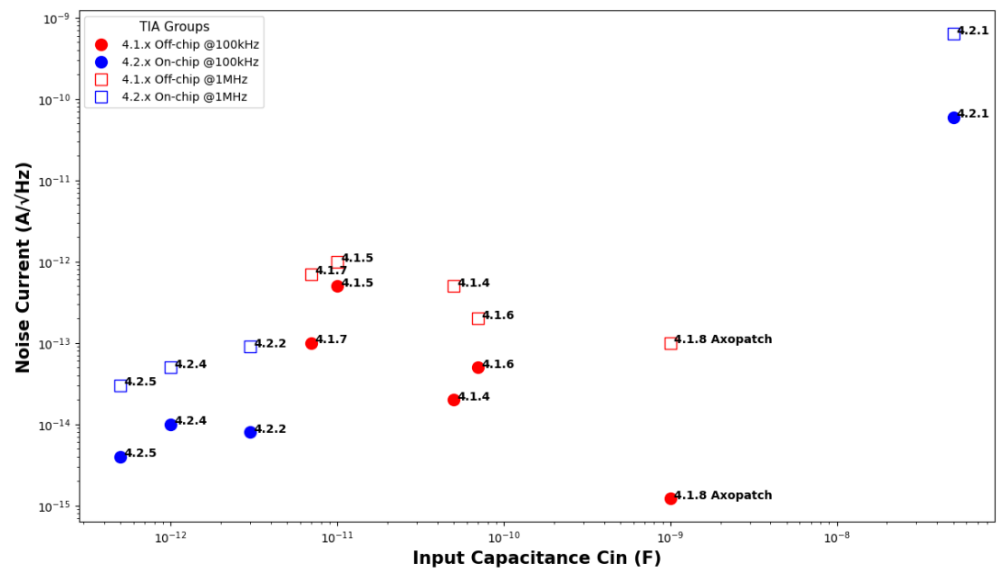

*Figure 21: Effect of Input Capacitance on Noise in On-Chip and Off-Chip TIAs at 100kHz & 1MHz*

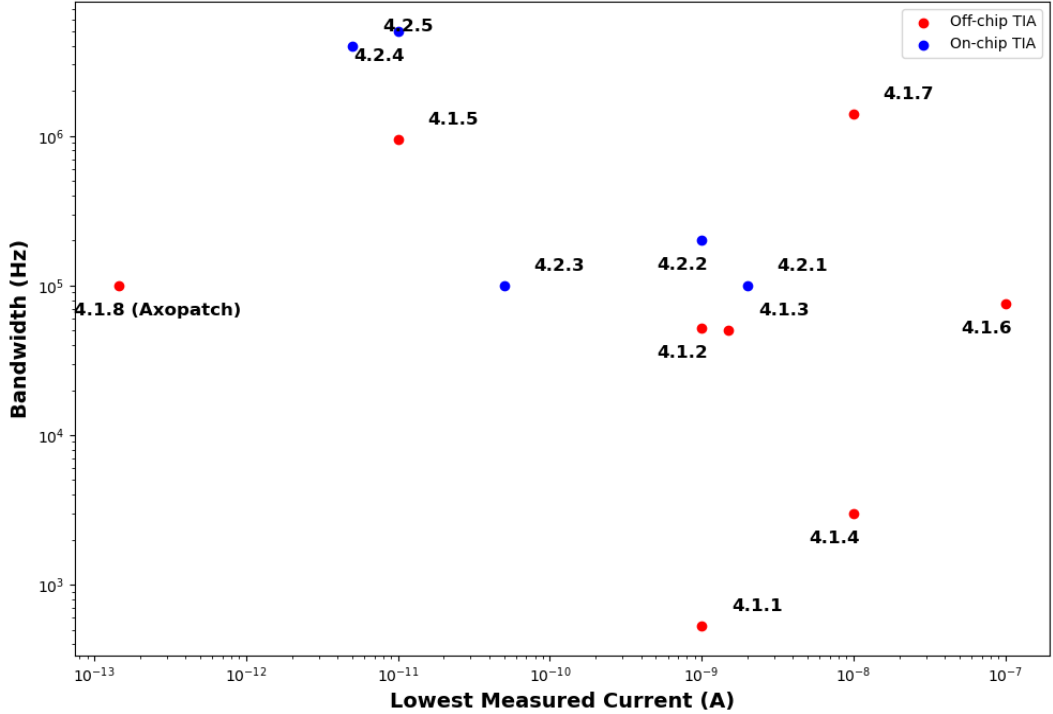

*Figure 22: Comparing Lowest measured Current and Bandwidth in On-Chip vs Off-Chip Models.*

Among the various design considerations for transimpedance amplifiers (TIAs), input parasitic capacitance emerges as a critical differentiator between on-chip and off-chip implementations. On-chip TIAs, realized through monolithic integration, offer significantly lower parasitic capacitance due to compact transistor geometries, short interconnects, and the absence of bond wires and PCB traces. The use of low K-dielectric materials and minimized junction areas further enhances their performance. In contrast, off-chip TIAs typically suffer from elevated input capacitance including op amp capacitance, cable capacitance, residual nanopore capacitance, and extended parasitic caused by extended PCB traces, solder joints, and discrete components. As input capacitance increases (4.2.1), noise levels rise due to greater interaction with the amplifier's voltage noise. Nanopore signals are required to operate at higher frequencies, where capacitive noise becomes a significant concern. This noise originates from the amplifier's input noise ($e_i$) and couples through the input parasitic capacitance ($C_{in}$), which is introduced by PCB traces, cables, and the amplifier itself. The power spectral density (PSD) of this noise is given by: $S_{I,cap} = (2\pi f C_{in})^2 e_i^2$. At higher frequencies, capacitive noise increases proportionally to the square of the frequency ($f^2$). As a result, the overall noise level rises significantly, which further degrades signal clarity and reduces the system's effective bandwidth. This distinction is especially crucial in nanopore sensing applications, where minimizing total input capacitance is essential for high bandwidth and low noise performance. While recent advances in microfabrication have significantly reduced membrane capacitance, the benefits can only be fully realized if amplifier parasitic capacitance is also minimized. On-chip application specific integrated circuit (ASIC) designs effectively address this challenge, enabling superior noise performance and signal integrity by reducing the total input capacitance and supporting high density integration for parallel sensing. Instead of attempting to reduce capacitance, as discussed in section 4.1.4, additional noise reduction techniques must be employed. While off-chip TIAs often struggle with noise due to parasitic capacitances, the Axopatch 200B (4.1.8) offers significantly lower noise than other off-shelf amplifiers. Its advanced design, integrated functionalities, and active cooling effectively minimize thermal fluctuations, ensuring superior signal fidelity and making it the preferred choice for high precision applications. From the analysis of the Current Range vs. Bandwidth graph, it is evident that the minimum current handling capacity of a Transimpedance Amplifier (TIA) is primarily influenced by board leakages and intrinsic noise characteristics. On-chip TIAs typically demonstrate lower leakage currents and reduced intrinsic noise levels compared to their off-chip counterparts, thus enabling effective handling of smaller current ranges. Conversely, off-chip TIAs generally experience higher leakage currents and increased noise, constraining their minimum current handling capabilities. However, certain off-chip TIAs, notably the Axopatch, deviate from this norm by employing advanced cooling and leakage subtraction methods, significantly reducing leakage and intrinsic noise. Such design enhancements, including specialized noise cancellation techniques as demonstrated in section 4.1.5, enable off-chip systems to handle smaller currents effectively. Ultimately, while the maximum current limit for both on-chip and off-chip TIAs remains dependent on the supply voltage. Furthermore, the bandwidth of TIAs is directly influenced by both the input capacitance ($C_{in}$) and the current magnitude. As capacitance increases or charge(current) decreases, the voltage across the capacitor according to the relationship $V_{out} = \frac{1}{C} \int i(t)dt$ decreases. Consequently, larger capacitances require more time to charge and discharge, thereby limiting the achievable bandwidth. Additionally, lower currents prolong the charging and discharging cycles, further reducing bandwidth. Thus, TIAs designed with smaller input capacitances and relatively higher current magnitudes, commonly found in on-chip implementations, typically achieve higher bandwidths. Nonetheless, certain off-chip designs, such as design 4.1.5, overcome these limitations through specialized noise cancellation techniques, thereby attaining bandwidth performance comparable to that of on-chip TIAs. In addition to noise performance and Current limitations, on-chip TIAs are more power efficient due to optimized routing and reduced interconnecting length. They also allow for the precise implementation of extremely small capacitances for feedback, critical for high frequency applications through techniques such as metal-insulator-metal (MIM), metal-oxide-semiconductor (MOS), and metal-oxide-metal (MOM) capacitor integration. Off-chip designs, on the other hand, face limitations imposed by PCB fabrication constraints and inherent stray capacitance, making it difficult to realize such small components with accuracy. Furthermore, leakage currents are more pronounced in off-chip designs due to PCB surface leakage, negatively impacting the accuracy of low current measurements. The discrete nature of off-chip components, larger physical layout, and higher exposure to external interference necessitate complex shielding and often compromise overall system performance. While off-chip TIAs retain certain advantages as discussed in Table I, such as ease of prototyping and flexibility in design iteration, their performance is inherently constrained by parasitic effects and physical limitations. In contrast, on-chip TIAs offer a comprehensive solution optimized for high speed, low noise, and energy efficient operation. Therefore, for precision driven applications like nanopore sensing, on-chip TIA architecture stands out as the superior choice, delivering enhanced scalability, reliability, and performance.

After thoroughly reviewing and discussing various Transimpedance Amplifier (TIA) designs, we summarize our findings through a concise comparison highlighting critical aspects such as Area, Signal to Noise Ratio (SNR), Power

Consumption, Parasitic, and Performance, Leakages. The following table compares the key characteristics of the Off-Chip and On-Chip implementations:

| Aspect | Area | SNR | Power Consumption | Parasitic Effects | Performance | Leakages |
|---|---|---|---|---|---|---|
| Off-Chip | Relatively Large Footprint | Moderate (Acceptable noise) | High (Increased power budget) | Significant (Higher parasitic effects) | Good (Adequate performance) | High (More leakage currents) |
| On-Chip | Compact (Reduced silicon area) | High (Superior noise performance) | Low (Improved power efficiency) | Minimal (Reduced Parasitic Influence) | Good (Enhanced, integrated performance) | Low (Reduced leakage currents) |